\documentclass[iop]{emulateapj}
\usepackage{natbib}
\usepackage{longtable}
\usepackage{changepage}

\begin{document}
\title{Building the Evryscope: Hardware Design and Performance}

\author{Jeffrey~K.~Ratzloff\altaffilmark{1}, Nicholas~M.~Law\altaffilmark{1}, Octavi~Fors\altaffilmark{1,2}, Henry~T.~Corbett\altaffilmark{1}, Ward~S.~Howard\altaffilmark{1}, \\ Daniel~del~Ser\altaffilmark{1,2}, Joshua Haislip\altaffilmark{1}}

\altaffiltext{1}{Department of Physics and Astronomy, University of North Carolina at Chapel Hill, Chapel Hill, NC 27599-3255, USA}
\altaffiltext{2}{Dept. de F{\'i}sica Qu{\`a}ntica i Astrof{\'i}sica, Institut de Ci{\`e}ncies del Cosmos (ICCUB), Universitat de Barcelona, IEEC-UB, Mart\'{\i} i Franqu{\`e}s 1, E08028 Barcelona, Spain}

\email[$\star$~E-mail:~]{jeff215@live.unc.edu}


\begin{abstract}
The Evryscope is a telescope array designed to open a new parameter space in optical astronomy, detecting short timescale events across extremely large sky areas simultaneously. The system consists of a 780 MPix 22-camera array with an 8150 sq. deg. field of view, 13" per pixel sampling, and the ability to detect objects down to $m_{g'}\simeq$16 in each 2 minute dark-sky exposure. The Evryscope, covering 18,400 sq.deg. with hours of high-cadence exposure time each night, is designed to find the rare events that require all-sky monitoring, including transiting exoplanets around exotic stars like white dwarfs and hot subdwarfs, stellar activity of all types within our galaxy, nearby supernovae, and other transient events such as gamma ray bursts and gravitational-wave electromagnetic counterparts. The system averages 5000 images per night with $\sim$300,000 sources per image, and to date has taken over 3.0M images, totalling 250TB of raw data. The resulting light curve database has light curves for 9.3M targets, averaging 32,600 epochs per target through 2018. This paper summarizes the hardware and performance of the Evryscope, including the lessons learned during telescope design, electronics design, a procedure for the precision polar alignment of mounts for Evryscope-like systems, robotic control and operations, and safety and performance-optimization systems. We measure the on-sky performance of the Evryscope, discuss its data-analysis pipelines, and present some example variable star and eclipsing binary discoveries from the telescope. We also discuss new discoveries of very rare objects including 2 hot subdwarf eclipsing binaries with late M-dwarf secondaries (HW~Vir systems), 2 white dwarf / hot subdwarf short-period binaries, and 4 hot subdwarf reflection binaries. We conclude with the status of our transit surveys, M-dwarf flare survey, and transient detection.

\end{abstract}


\section{Introduction}
Astronomical surveys searching for time-variable objects and events typically observe few-degree-wide fields repeatedly, use large apertures to achieve deep imaging, and tile their observations across the sky. The resulting survey, such as the Palomar Transient Factory \citep{2009PASP..121.1395L}, Pan-STARRS \citep{2010SPIE..7733..123K, 2012ApJ...750...99T}, SkyMapper \citep{2007PASA...24....1K}, ATLAS \citep{2011PASP..123...58T}, CRTS \citep{2011arXiv1102.5004D}, ZTF \citep{2018arXiv180210218B}, and many others, is necessarily optimized for events such as supernovae that occur on day-or longer timescales. These surveys are not sensitive to the very diverse class of shorter-timescale objects, including transiting exoplanets, young stellar variability, eclipsing binaries, microlensing planet events, gamma ray bursts, young supernovae, and other exotic transients, which are currently only studied with individual telescopes continuously monitoring relatively small fields of view, or groups thereof. Short-timescale surveys including HAT \citep{2004PASP..116..266B}, SuperWASP \citep{2006PASP..118.1407P}, KELT \citep{2007PASP..119..923P}, and many others observe dedicated sky areas to reach very fast cadence and good sensitivity, but at the expense of all sky coverage. The Evryscope is designed to reach bright but rare events by optimizing for shorter-timescale observations with continuous all sky coverage continued for many years.

\begin{figure}[tbp]
\includegraphics[width=1.0\columnwidth]{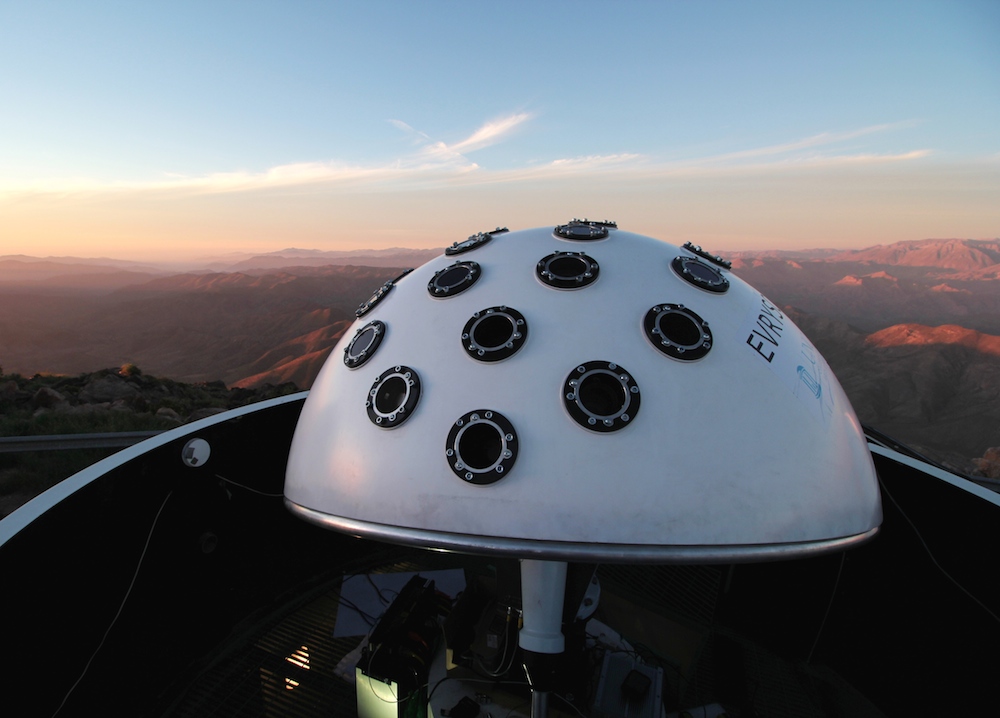}
\caption{The Evryscope, a two-dozen-camera array mounted into a 6 ft-diameter hemisphere, deployed at the CTIO observatory.}
\label{fig:overview_fig}
\end{figure}

The Evryscope (Figure \ref{fig:overview_fig}) uses an array of 22 telescopes to cover the Southern sky down to an airmass of $\approx$2.0 in each exposure. The system averages 5000 images per night with $\sim$300,000 sources per image. The Evryscope features mass-produced compact CCD cameras and lenses, and a novel camera mounting scheme to make a reliable, low-cost 0.8 gigapixel robotic telescope. We built the Evryscope at UNC Chapel Hill in early 2015 and deployed it to CTIO in Chile in May 2015. The system has collected data continuously since first light in May 2015. As of March 2019, we have taken over 3.0M images resulting in 250TB of raw data. The resulting light curve database has light curves for 9.3M targets down to $m_{g}$=15 (and fainter for selected targets), averaging 32,600 epochs per target through 2018.

The Evryscope mounts an array of individual telescopes into a single hemispherical enclosure (the "mushroom"). The array of cameras defines an overlapping grid in the sky providing continuous coverage of 8,150 square degrees. The camera array is mounted onto an equatorial mount which rotates the mushroom to track the sky with every camera simultaneously for 2 hours, before "ratcheting" back and starting tracking again on the next sky area (Figure \ref{fig:cutaway}). Each of the telescopes has three-hundred-square-degree fields of view, 28.8 megapixels, and a 6.1cm aperture. The Evryscope allows the detection and monitoring of objects and events as faint as $m_{g'}$=16.5 in few-minute exposures ($m_{g'}$=15-16 under typical sky conditions) and as faint as $m_{g'}$=19 after co-adding. The telescope specifications are given in Table \ref{tab:overview_tab}. 

The Evryscope has already contributed to a wide variety of science cases, ranging from precision studies of single targets \citep{2018AJ....156..120T,2019AJ....157...97K} and (Ratzloff et al., in prep), to statistical studies of stellar activity (Howard et al., in prep), variable star discoveries (Ratzloff et al., submitted), hot subdwarf / white dwarf short-period binary discoveries (Ratzloff et al. in prep.), and transient discovery and followup \citep{2041-8205-860-2-L30, {2018ATel11467....1C}}. In this paper we, in addition to describing the Evryscope hardware, describe some of the first Evryscope discoveries from general stellar searches. A previous paper \citep{2015PASP..127..234L} describes the detailed Evryscope science cases. Subsequent papers will describe the data analysis pipelines in detail. 

This paper is organized as follows: in \S~\ref{section_design} we explain the Evryscope system, design, and primary components. In \S~\ref{section_performance} we describe the on sky performance. \S~\ref{section_results} describes the transit detection methods, and shows example light curves and select first discoveries. In \S~\ref{section_summary} we conclude.


\section{System design} \label{section_design}

\begin{table*}[ht]
\caption{The specifications of the Evryscope}
\centering
\begin{tabular}{l l}
\hline
Hardware & Description \\
\hline
Telescope mounts & 27 (22 populated); shared equatorial mount \\
Telescope glass & 61mm Rokinon F1.4 lenses\\
Mechanical mounting & Fiberglass dome with aluminium supports \\
Detectors &  28.8MPix KAI29050 interline-transfer CCDs \\
 & 7e- readout noise at 4s readout time \\
 & $\approx$50\% QE @500nm; 20,000 e- full-well capacity \\
 Field of view (Measured on sky) & 8150 sq. deg. total (excluding $\approx$10\% overlaps)\\
Sky coverage per night & 18,400 sq. deg. (2-10 hours per night coverage) \\
Total detector size & 780 MPix \\
Sampling & 13" /pixel \\
Observing strategy & Track for 2 hours; reset and repeat \\
Data storage &  All data recorded for long-term analysis \\
 & $\sim$50TB / year after all overheads \\
\hline
Performance & Description \\
\hline
PSF 50\% enclosed-energy diameter & 2 pixels in central 2/3 of FoV; 2-4 pixels in outer 1/3 \\
Exposure time & 120s \\
Limiting magnitude & $m_{g'}$=16.0 (3-sigma; 120s exposure) \\
Photometric performance & 1\% photometry on $m_{g'}<$12 stars every 2 minutes \\
 & 6 \% photometry on $m_{g'}$=13.5 every 2 minutes \\
 & 10 \% photometry on $m_{g'}$=15.0 every 2 minutes \\
\hline
\end{tabular}
\label{tab:overview_tab}
\end{table*} 

\subsection{Science requirements}
The Evryscope's science requirements were based on a study of the science possibilities for an all-sky telescope with an Evryscope-like design, detailed in \cite{2015PASP..127..234L} and summarized in Table \ref{tab:science_cases_tab}. With eighteen major science cases for the system, each of which having somewhat different needs, the setting of exact requirements was challenging. To constrain the design space and allow choices to be made, we settled on three simple requirements: a field of view around 8,000 square degrees, a 3-sigma limiting magnitude of $m_{g'}\simeq$16, a pixel scale sufficient to avoid crowding for 90\% of sources above a galactic latitude of $15^\circ$, photometric precision better than 1\% for bright stars, and the ability to co-add images to increase the target depth.

\begin{table*}[ht]
\caption{The Evryscope science cases}
\centering
\begin{tabular}{l l}
\hline
Field & Description\\
\hline
Exoplanets & White-dwarf transits \& debris disks \\
           & Hot-subdwarf transits \& debris disks \\
           & Habitability-affecting superflares \\
           & Eclipse timing exoplanet detections \\
           & Confirmation of TESS single-giant-planet-transit events \\
           & Long-period rocky exoplanets transiting M-dwarf stars \\
Stellar astrophysics  & Low-mass-star rotation and activity\\
                      & Long-period eclipsing binaries for mass-radius relations\\
                      & Young-star activity and multiplicity\\
                      & Star-planet activity interactions\\
                      & Interacting binary outbursts\\
                      & Long-period dust dips\\
Transients & Gravitational-wave electromagnetic counterparts\\
           & Microlensing exoplanet detection\\
           & Galactic nova events\\
           & Nearby, young supernovae\\
           & Gamma-ray burst counterparts\\
           & Fast-radio-burst counterparts\\

\hline
\end{tabular}
\label{tab:science_cases_tab}
\end{table*}

\subsection{Overall design}
\begin{figure}[tbp]
\includegraphics[width=1.0\columnwidth]{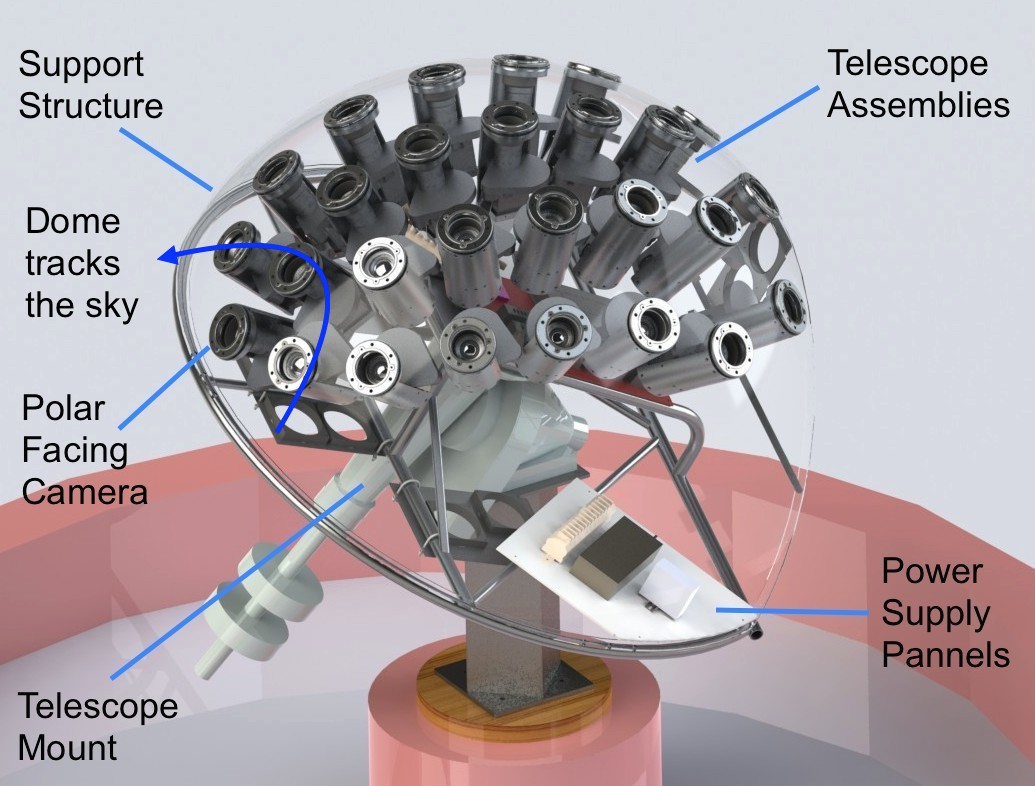}
\caption{Cutaway rendering of the Evryscope showing the telescope mount, camera locations, and primary instrument components.}
\label{fig:cutaway}
\end{figure}

Starting with the general plan of an array of telescopes mounted together, we evaluated several concepts for the overall system design, including a flat tracking platform with each camera bolted to it, adjustable trusswork supporting each camera, and a spherical-shape rotated around its polar axis \citep{2012SPIE.8444E..5CL}. We settled on a hemispherical dome mounted on an equatorial mount (the ``mushroom''). This offered two advantages: the camera support structure could be a single piece with no per-camera adjustment or alignment required, and the tracking mount, the single moving main structure and therefore critical to reliability, could be a single off-the-shelf system. We summarize our overall design in Figure \ref{fig:cutaway}.

\subsection{Camera array design}
An Evryscope-type array telescope design has an enormous range of possible design choices. The choice of CCD array size must be traded-off against the choice of lens, the point-spread-function (PSF) quality available over the chosen array size, the pixel scale resulting from a particular lens/CCD combination, and more subtle factors like vignetting and angular quantum efficiency.  With the CCD detectors being the driving cost, the science requirement flowdown to the technical requirements was informed by a hardware-budget target of $\approx$\$300k.

\subsubsection{Lens and CCD choice}
With dozens of lenses and CCD-arrays available from a multitude of manufacturers, we performed a comprehensive trade study of the possible lens/CCD combinations. The pixel scale was set by the anti-crowding science requirement to be smaller than 20\arcsec, and we set the field of view to 8,000 square degrees. With those parameters fixed, we evaluated each lens/CCD combination based on the SNR that could be achieved all-sky on a $m_{g'}$=16 source.  The SNR calculations included the likely PSFs and vignetting generated by each lens/CCD combination, the expected sky background and source photon noise contributions, the detector characteristics, and many other factors, and most lens/CCD combinations were not able to achieve the required SNR because of one of those factors. 

We elected to limit our CCD selections to interline-transfer chips which have electronic shutters. Our prototype systems \citep{2013AJ....145...58L} both suffered mechanical shutter failures during their arctic deployments, with the achieved number of error-free exposures being just over one-tenth the specification. Although the failures were correctable by individually adjusting the tension of internal springs every few months, this is untenable in a robotic system with dozens of cameras. The use of electronic shutters effectively eliminates this failure mode.

The trade study resulted in a single workable choice for lens/CCD combination: a Rokinon 85mm F/1.4 lens combined with a KAI29050 CCD array. All other combinations resulted in unacceptably-low SNR or budgets factors-of-several times larger than our target amount.  The KAI29050 array had a particular advantage in its rectangular format: most photographic lenses have rapid fall-offs in PSF quality towards the edges of the frame, and square arrays can therefore have poor image quality in the corners \citep{2013AJ....145...58L}. Compared to a square format, a rectangular array trades off highly-off-axis image area at the corners for less-off-axis area at the left and right edges of the array, and thus has more uniform PSFs across the image than a square CCD with equivalent area. Based on our positive experience with previous similar cameras, we elected to use thermoelectrically-cooled Finger Lakes Instrumentation ML29050 units.

\subsubsection{Camera position optimization and system field of view}
We next built a metric to optimize the camera positions in the array. Each camera produces a rectangular field on the sky, with a large enough field of view that spherical geometry must be taken into account for even simple sky-area calculations. We designed the camera array positions to 1) optimally tile over the above-airmass-two field of view; and 2) avoid large areas of overlap between cameras; 3) retain a few-degree overlap between each camera to constrain systematics. 

\begin{figure}[tbp]
\includegraphics[width=1.0\columnwidth]{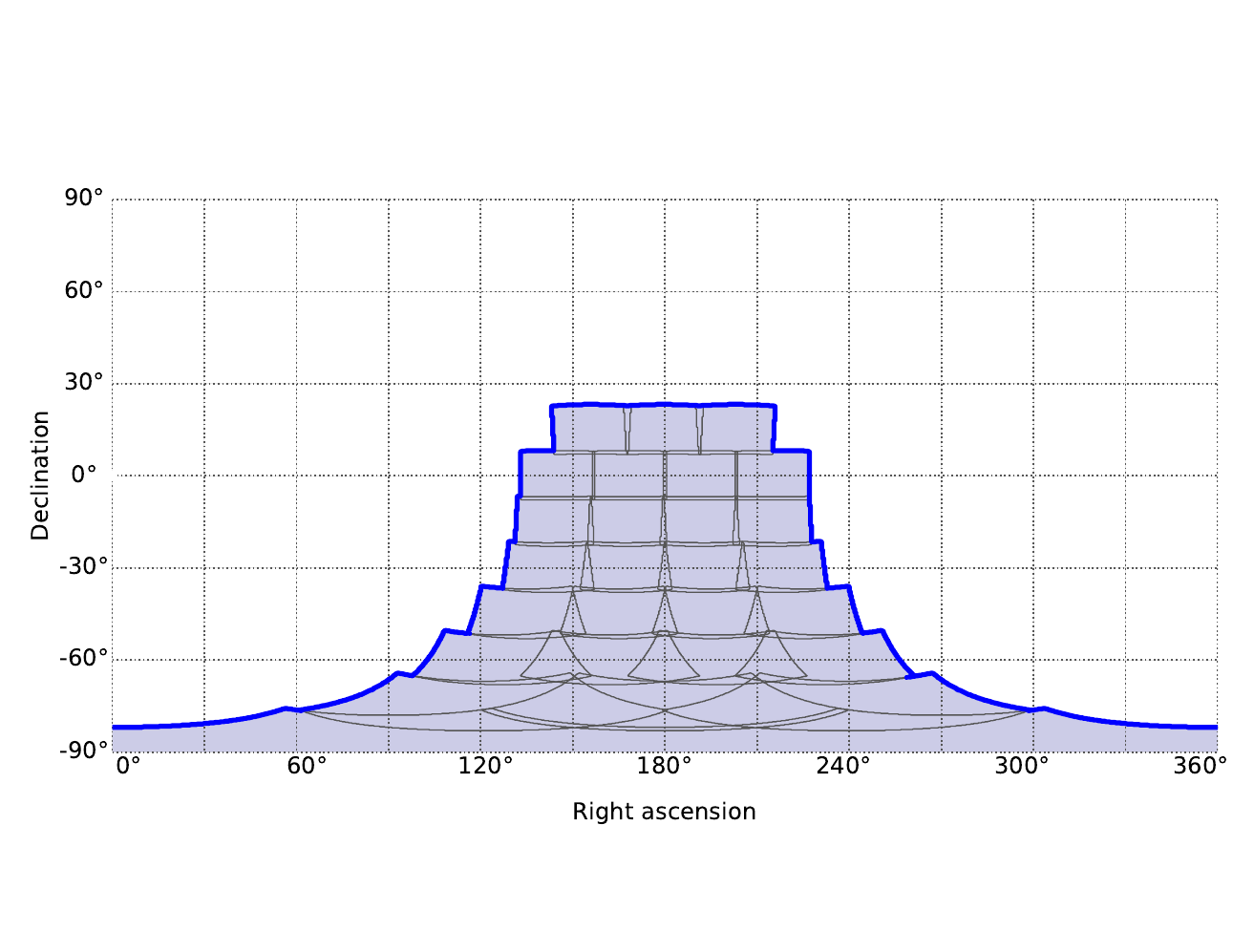}
\caption{The Evryscope camera placement when deployed at the CTIO observatory (some of the Northern camera spots are currently unpopulated).}
\label{fig:camera_placement}
\end{figure}

We designed a code to project the field of view of each camera onto the sky, taking spherical geometry into account. The code then divides the sky into patches approximately $0.3^{\circ}$ across, counts the number of cameras pointed at each patch, and measures the total sky area and overlap areas covered between different combinations of cameras. Starting with a simple arrangement of cameras divided into rows of declination, we then varied the position of each camera in the array using an annealed downhill-simplex algorithm, optimizing for overlap and covered sky area \citep{2016SPIE.9906E..1ML}. The optimization converged on an arrangement very similar to the input declination-separated grid of cameras; other camera arrangements we explored did not produce significantly better performance metrics. For ease of fabrication we used the simple declination-separated grid to place the cameras, with spacing parameters inherited from the fully-optimized solution (Figure \ref{fig:camera_placement}).

\begin{figure}[tbp]
\includegraphics[width=1.0\columnwidth]{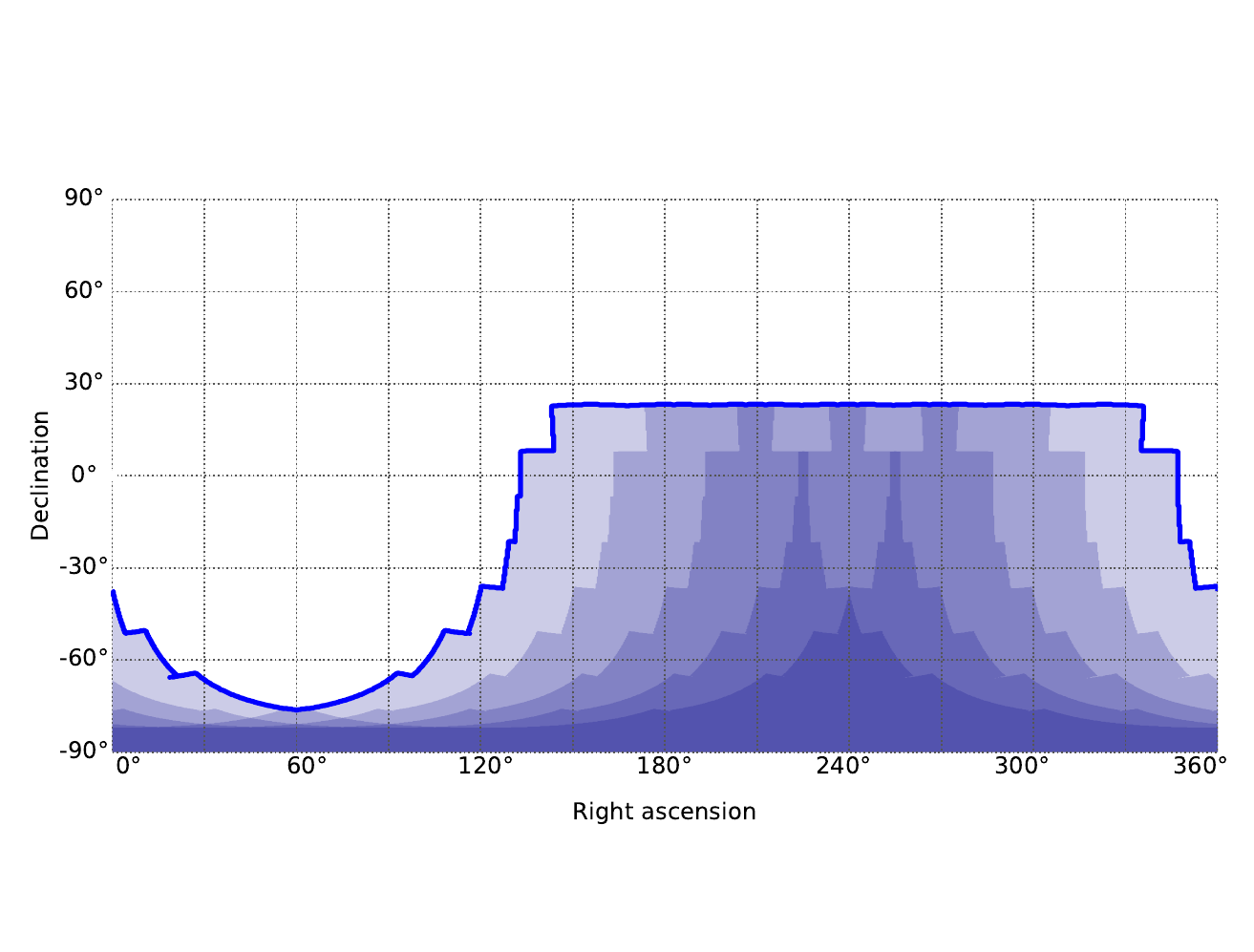}
\caption{The 18,000 square degree coverage of the system over a single night. The depth of coloration corresponds to the number of two-hour ratchets covering each part of the sky; each ratchet includes sixty two-minute epochs.}
\label{fig:sky_coverage}
\end{figure}

Each camera assembly rotates in a circular arc around the pole facing camera as the mushroom tracks the sky. Over the course of a typical night the system covers $\approx18,000$ sq. deg. (Figure \ref{fig:sky_coverage}), with each part of the sky being observed at two-minute cadence for 4-10 hours per night.

Each CCD is orientated so that its long axis (designated as the x-axis) is tangential to this arc; this ensures the objects in each image remain in a constant orientation throughout the night. There are seven rows, with the cameras in each row sharing the same pointing declination, equidistant from the pole camera. The camera mounting flanges (and therefore the CCDs) are normal to the surface of the mushroom dome, which ensures that the cameras are pointed in the proper direction without manual alignment being necessary. We designed the mushroom to be capable of supporting 27 telescopes; at CTIO 24 are Southern hemisphere facing and three cover positive declinations. The number of operational cameras has varied slightly during the course of the project: 22 or 23 cameras have been operational in 2015-2017, with another camera reserved for testing. We plan to fill in all available slots in the near future.

\subsection{Telescope structure, tracking and image quality optimization} \label{telescope_structure}
Mechanically, the Evryscope consists of an array of cameras mounted into a hemisphere (the mushroom), which in turn is mounted onto a German-equatorial mount which keeps all the cameras tracking.

\subsubsection{Camera hardware units} 
The camera hardware units fix the cameras to the mushroom, provide mechanical support of the components, and a mount for a protective window. The camera mounts have three primary constraints on their design: flexure limits, size and weight. Although atmospheric refraction precludes keeping each star on the same pixel while tracking \citep{2015PASP..127..234L}, we designed the camera mounts to not contribute any extra drift throughout the Evryscope's range of motion, requiring the relative camera mount flexure to be less than 13\arcsec. The size of the mushroom was set to a 6-foot diameter by our target dome, and this set the packing requirements for the cameras.  Since there are two dozen camera mounts with relatively heavy CCD units, they and the systems they contain are the primary drivers of the weight of the system. A trade study of available mounts suggested that significant cost savings were possible if the total mushroom weight could be kept below 400 lbs.

\begin{figure}[tbp]
\centering
\includegraphics[width=1.0\columnwidth]{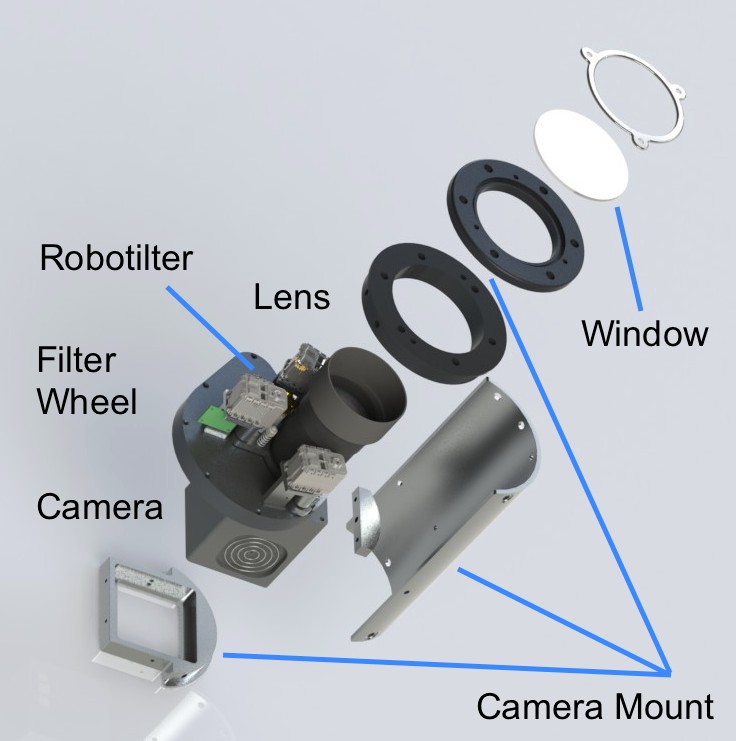}
\caption{The Evryscope unit camera assembly}
\label{fig:camera_mount}
\end{figure}

We used 3D modeling to test several hardware unit designs, with the goal to minimize weight, flexure, and complexity. The final version (Figure \ref{fig:camera_mount}) features interlocking sections for added rigidity, weighs less than 4 lbs (supporting imaging hardware which weighs 8.0 lbs), and provides a maximum differential flexure of less than 10 arcsec. The maximum flexure in the vertical orientation is $\approx$.02 mm and over the course of a telescope ratchet the differential movement due to the changing camera orientation is well within our 1 pixel goal. The camera mounts are interchangeable, have locator pins to easily place the cameras into the proper orientation in the mushroom, and perform equally well in flexure for all cameras regardless of the declination row (which have considerably different gravitational vectors). 

Each mount has an outer window to protect the lenses and electronics from dust, water, and other possible contaminants, enabling easy cleaning as well a providing a backup to the observatory dome. The high transmission (over 96\% in the visible range) optical window is mounted on a soft o-ring with a stainless steel retaining ring, and allows for easy cleaning of dust during maintenance.  

Interline transfer CCDs cannot take darks without extra mechanical shutters, so we elected to use a filter wheel with a blocked position to allow calibrations to be taken. The Finger Lakes Instrumentation CFW-5-1 filter wheels also provide a sunshield (\S~\ref{subsub_sunshield}) and science filter changing capability.

\subsubsection{Precision lens/CCD alignment systems (``Robotilters'')} 

Camera lenses are used on SuperWASP \citep{2006PASP..118.1407P}, HAT \citep{2004PASP..116..266B}, KELT \citep{2007PASP..119..923P}, XO \citep{2005PASP..117..783M}, and other transiting exoplanet surveys to reach as much as 1000 square degree fields of view. Other surveys types such as the ASAS-SN (supernova) \citep{Shappee2014}, Pi of the Ski (gamma ray bursts) \citep{2013A&A...551A.119P}, Fly's Eye (asteroid detection) \citep{2013ASPC..475..369C}, and HATPI \footnote{\url{https://hatpi.org}} also use camera lenses to reach wide sky coverage. These types of wide field surveys and many others including the Evryscope are susceptible to image quality tilt and focus challenges.   Even a slight misalignment between the optics and the CCD causes a tilt which results in an unacceptable increase in size of the PSF FWHM towards the edges and corners of the image. For the Evryscope, the very wide field of view (380 sq. deg.), fast F\# of each lens and the small 5.5$\mu m$ pixels exaggerate this effect. While the machining tolerances (+/-.005 inch in most cases) and the assembly tolerances of the mass produced lenses, adapters, filter wheels, and CCD assemblies is reasonable for their standard usages, it is not precise enough to achieve the absence of tilt required for the needed Evryscope image quality.

\begin{figure}[tbp]
\includegraphics[width=1.0\columnwidth]{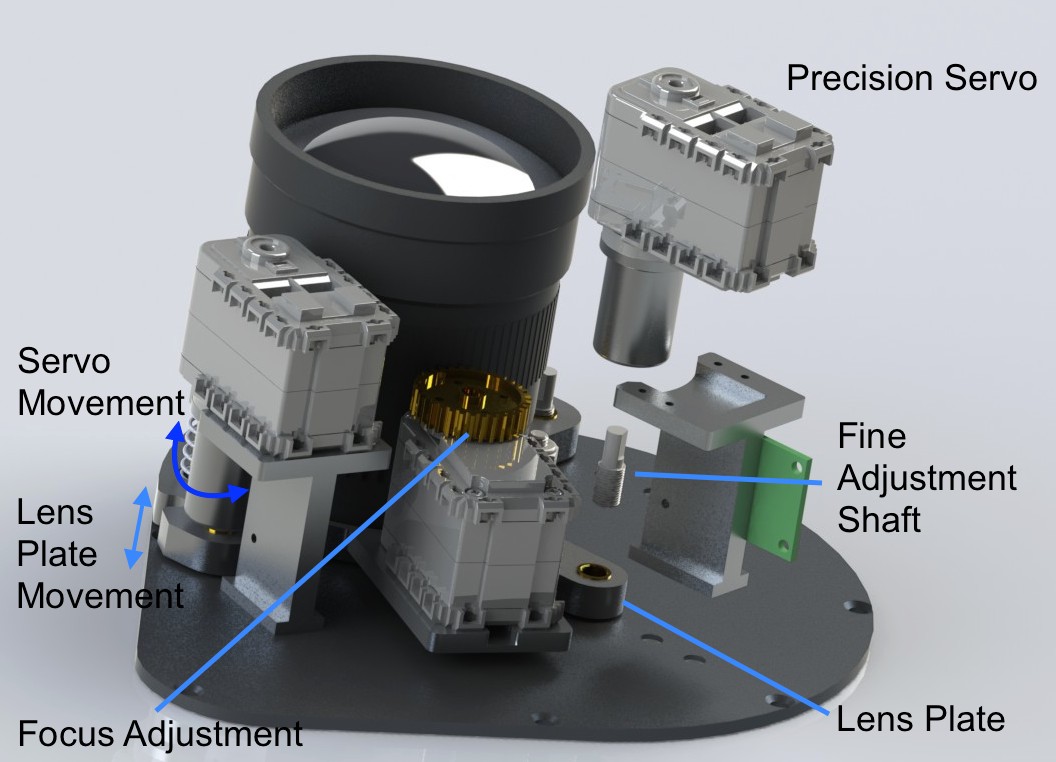}
\caption{The Robotilter automated tilt/alignment/focus system}
\label{fig:robotilter}
\end{figure}

We designed a robotic tilt adjustment mechanism (Figure \ref{fig:robotilter}) to address those challenges, with the ability to remotely and precisely re-align the camera assemblies. The Robotilter (Figure \ref{fig:robotilter}) uses three precision servos controllable to within 4 degree steps coupled to an 80 thread per inch adjuster to move the lens position relative to the CCD. This allows adjustment of the tilt as well as the lens/CCD separation in increments as fine as .003 inch. The design uses specialized flexible shaft couplings to prevent binding and tension springs to hold the lens accurately in place. The assembly mounts to the top plate of the filter wheel to avoid costly re-configuring of the existing filter wheel, CCD, or camera mount. A separate servo independently adjusts the lens focus position to compensate for tolerance differences due to temperature changes throughout the year. The Robotilters were installed in November 2015 and the cameras were aligned remotely in early 2016; the installation of the Robotilters was the final step in commissioning the system. The Robotilters and resulting image improvements will be described in detail in an upcoming technical paper (Ratzloff et al. in prep).

\subsubsection{Mushroom structure and wind shake}
The camera support structure (the mushroom, Figure \ref{fig:overview_fig}) needs to provide the same limited flexure as the camera mounts, while also bearing the 400lbs load of up to 27 camera assemblies and related components. We chose a molded fiberglass hemisphere with support ribs along the bottom and back for extra strength and rigidity, and a sturdy mounting point. The material is hand-laid cloth weave fiberglass, providing light weight and minimal flexure with excellent durability. The mushroom also features reinforced and precison-located inner and outer camera-mount flanges to provide accurate and secure mounting points. The camera flanges are normal to the surface, and the holes are CNC cut into the mushroom to ensure the precise location necessary to achieve the desired field coverage without holes or excessive overlap. The manufacturing tolerances are .020 inch on the hole locations, and based upon this the camera alignment is fixed normal to the mushroom surface and the long side CCD is perpendicular to the rotation axis. Our 3D model simulation predicts that despite the close packing of the cameras and considerable weight, the stress is mostly compression and results in absolute movements on the scale of .02 mm. Differential camera movements over the tracking cycle are on the order of microns ensuring accurate camera pointing. On-sky pointing accuracy is well within the simulated performance.

The hemispherical shape of the mushroom, along with the placement of the instrument so that the dome leafs in the open position are slightly higher than the mushroom base, help make the Evryscope resilient to wind shake. The system is able to operate in 30 mph winds without a measurable change in image quality.

\subsubsection{Tracking mount}
\begin{figure}[tbp]
\epsscale{1}
\includegraphics[width=1.0\columnwidth]{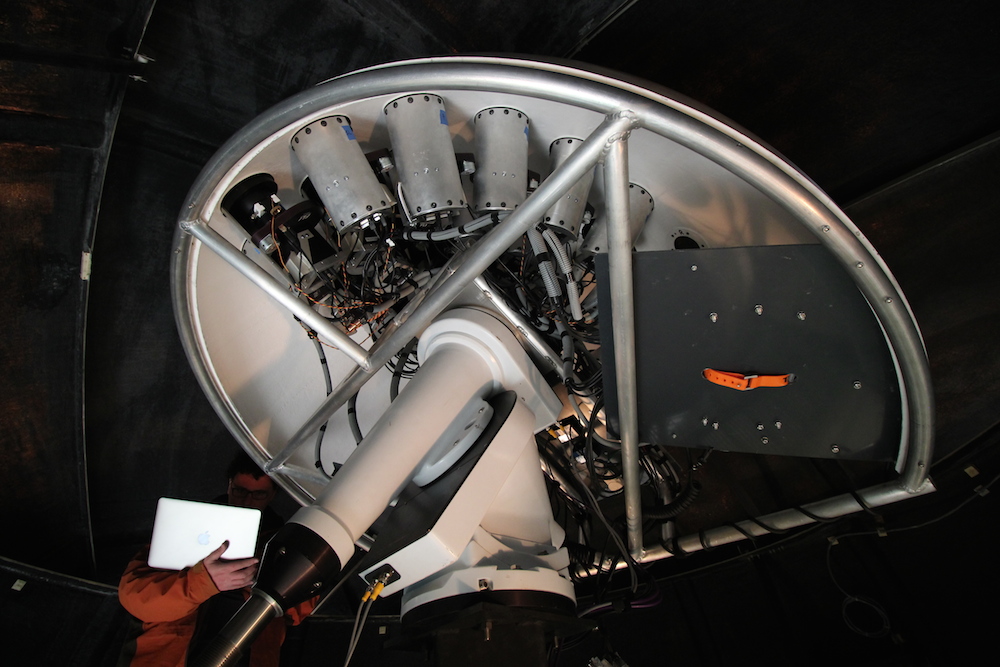}
\caption{The Mathis German Equatorial mount, the tubular base structure, and the mounted mushroom - showing the instrument inset, mass alignment, and camera accessibility.}
\label{fig:mushroom_mount}
\end{figure}

The base structure (Figure \ref{fig:mushroom_mount}) attaches the mushroom to the Mathis 750 mount, via a mount plate attached to the tracking mount and a structure which transfers the mechanical load from the mushroom fiberglass. We tested several design ideas via finite element analysis and found a reinforced round tubing design to be most effective. Using aluminum tubing, we reduced the weight in half from a similar design made of steel and kept the total flexure within requirements. The differential camera displacement of the mounting base throughout the telescope tracking is on the order of microns, and combined with the mushroom and camera mount flexure is simulated to be within our total goal of 1 pixel, with comparable performance measured on-sky.

The proper location of the center of mass is critical to reliable telescope mount operation. We inset the mount plate significantly into the mushroom so that the effective lever arm of the Evryscope cameras is minimized (Figure \ref{fig:mushroom_mount}). The center of mass is only 10 inches from the mount plate, which greatly reduces the load on the telescope mount compared to simpler designs. The base structure positions the Evryscope so that the center of mass in the mounted position is directly over the telescope mount axis center, further reducing stress on the telescope mount and easing the balancing of the instrument.

The polar alignment of the mount is critical to the tracking performance of the system. Because the system's field of view is such a large fraction of the sky, conventional pointing models cannot be used, because they optimize the performance on one part of the sky by reducing performance on other parts of the sky. For this reason we developed a precision polar alignment procedure specifically for Evryscope-like instruments (\S\ref{sec:polar_align}).

On sky performance confirms the predictions of the flexure and center of mass simulations. The camera pointing is accurate within a tenth of a degree, providing the proper field of view overlaps without gaps (except for one initial, now corrected, misalignment caused by a contaminated bolt thread). The camera orientations remain constant throughout sky tracking. The telescope mount tracks the sky consistently without stalling or shifting, and we conclude that the total flexure is very close to the 1 pixel goal. 

\subsubsection{Dome}
The Evryscope is located in an AstroHaven clamshell dome originally built for the PROMPT network of telescopes \citep{2005NCimC..28..767R}. The dome had already been used for routine long-term operation, and no mechanical changes beyond a custom pier structure were necessary for the Evryscope deployment. Careful electrical design was necessary, however; the large dome opening/closing motors can induce strong transients onto power and potentially signal lines from the dome. To avoid possible interference or even damage, we separated the dome electrical systems on a separate UPS system. A Raspberry-Pi single-board computer runs the dome-control daemon and communicates with the rest of the system via an electrically-isolated ethernet connection; there are no other direct electrical links between the Evryscope and the dome.

\subsubsection{Observatory site \& weather-related design}
\begin{figure}[tbp]
\epsscale{1}
\includegraphics[width=1.0\columnwidth]{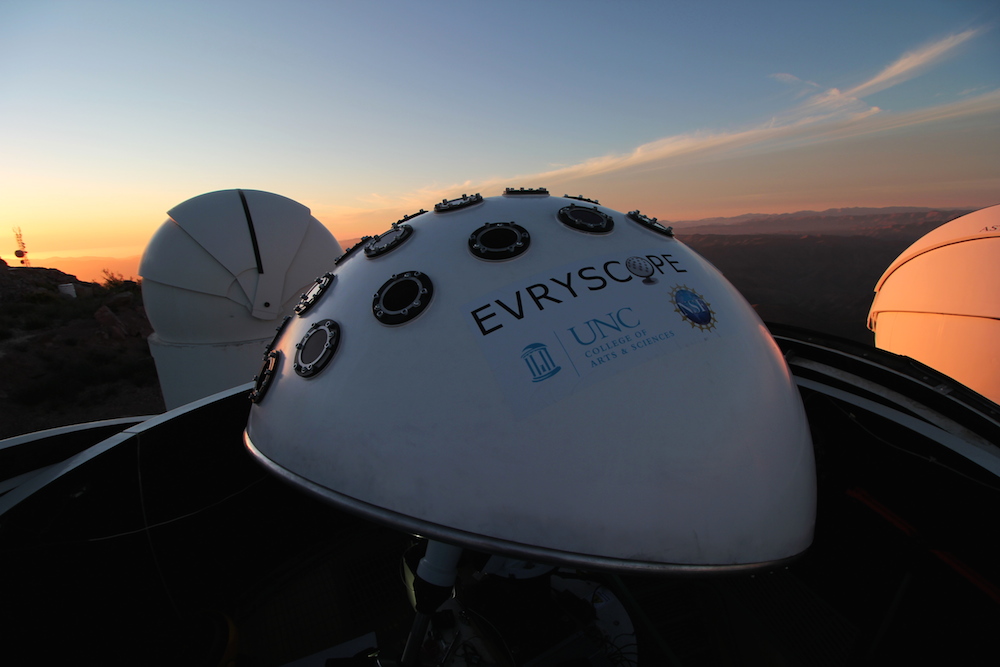}
\caption{The Evryscope in PROMPT dome 4.}
\label{fig:prompt}
\end{figure}

The Evryscope is deployed at CTIO in Chile in PROMPT \citep{2005NCimC..28..767R} dome 4 (Figure \ref{fig:prompt}). The site was chosen for the large number of usable nights ($>$ 320 per year), dark sky conditions ($m_{v}$ = 21.8 moonless night background average), and Southern sky visibility. UNC affiliated hardware and support synergies, especially the PROMPT Program, were also advantageous.

The dome and observatory site introduced several design constraints: 1) a maximum power consumption of 15A/120V; 2) operation with a relatively small internet bandwidth that precludes the realtime off-site transport of data; 3) the potential for lightning strikes and earthquakes (\S~\ref{system_robust}); 4) potential external temperature ranges of $-15^{\circ}$C to $+25^{\circ}$C; and 5) extremely dry conditions.
	    
The low end of the temperature range is outside that which most off-the-shelf electronics are rated for. Wherever possible we purchased industrial components rated for low-temperature operation (typically $-20^{\circ}$C). In some cases we tested and used off-the-shelf consumer electronics (for example, Raspberry-Pi single-board computers); testing was performed in fridge-freezer units under a range of relative humidity (see \cite{2013AJ....145...58L, 2016SPIE.9906E..1ML} for testing details).
	    
The potential for extremely dry weather spells required careful electronic and mechanical design. For example, Nylon becomes brittle under extremely dry conditions \citep{1989.060090206}; this can cause failures in cable insulation and zip-tie-type harnesses in a matter of months, leading to possible short circuits or mechanical interference between cables and moving parts. The static electricity discharges prevalent in dry conditions can cause electronic failures, especially while personnel are maintaining the system. Many power supplies and similar units are rated only to 20\% relative humidity, while the CTIO site can regularly reach low-single-digit humidity. We mitigated these concerns by using only plastics, connectors, and electronics rated for long-term extremely dry conditions. All metal components are grounded, with isolators used to avoid ground loop conditions, and we take operational steps to ground personnel before working on the system.

\subsection{Electrical and electronic design}
\begin{figure}[tbp]
\includegraphics[width=1.0\columnwidth]{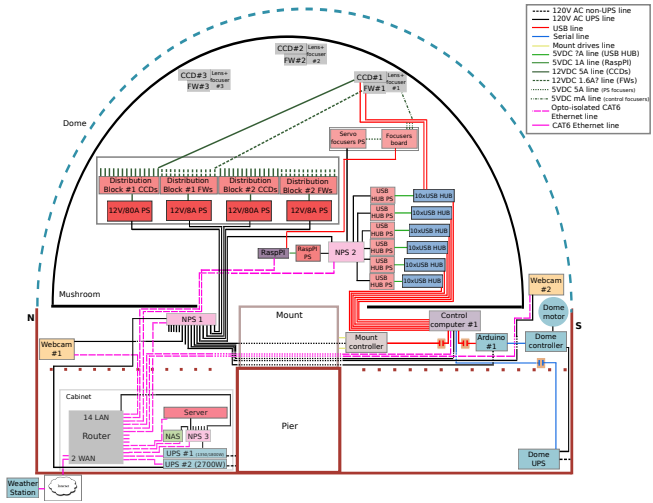}
\caption{The Evryscope Wiring Diagram.}
\label{fig:wiring}
\end{figure}

The Evryscope mushroom contains over 600ft of cabling, with further ancillary systems located outside the main telescope body. Figure \ref{fig:wiring} shows an overview of the power and data paths within the dome.
    
\subsubsection{Power distribution}
The Evryscope cameras together require a maximum of $\approx170A$ of 12V power; the ancillary systems with the mushroom (Robotilters, filter wheels, fans, USB hubs, etc.) together require a further $\approx$20A of 12V power. The AWG-1 (quarter-inch-diameter) cables required to safely carry the required 200A into the mushroom would be bulky and inflexible, and risky if frayed or overheated. Powering each camera from its own 12V supply would lead to a very bulky and heavy power distribution system, beyond the load capacity of the mushroom mount. For those reasons we elected to send 120V AC power into the mushroom over a single flexible small-diameter cable, and use two 120A-capable 12V power supplies to power the main camera systems. We deliberately overspecified the power supplies to reduce the need for active cooling and the associated vibrations. Ancillary systems are powered from their own smaller 12V power supplies, with Digital Loggers Network Power Switches allowing computer-controlled switching of each component. Although it has proven reliable, this setup resulted in over 600ft of cabling inside the main mushroom, because each camera has six separate cables going into it (3 power, 3 data). These cables are heavy and impede airflow; the Northern Evryscope, currently under commissioning, has relay and control systems built into each camera to reduce the number of required cables to two per camera.

\begin{figure}[tbp]
\includegraphics[width=1.0\columnwidth]{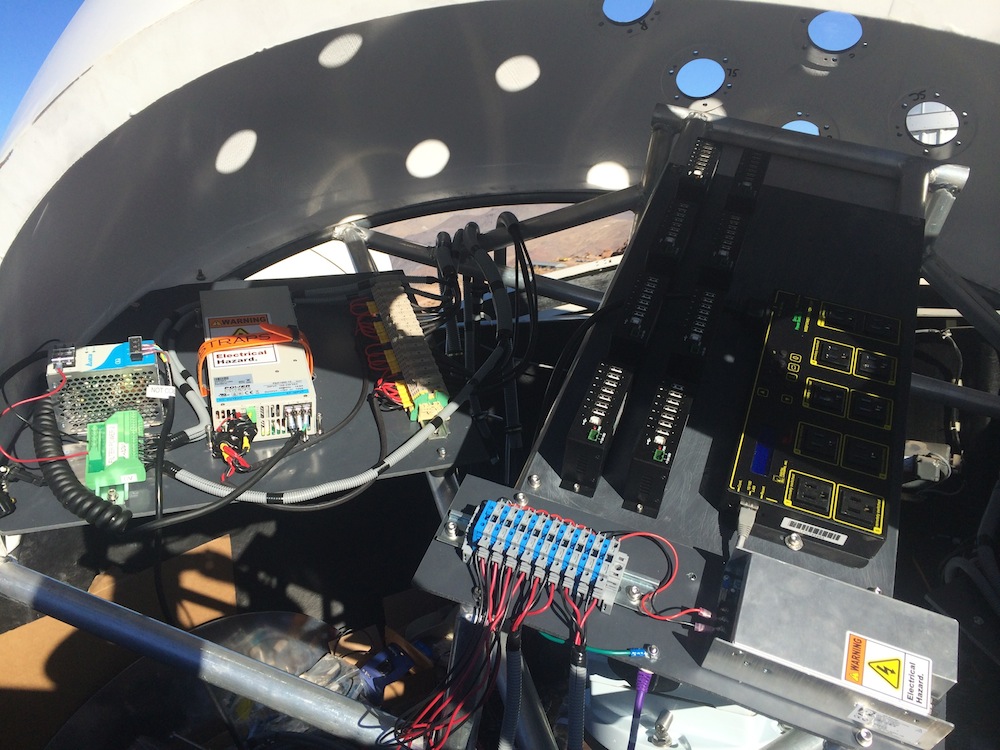}
\caption{The power supply panels; left is the camera and filter wheel power/distribution and the right are the USB hubs and NPS.}
\label{fig:panels}
\end{figure}

The two 120V input / 12V 80A output power supplies are mounted on panels attached to the wings of the base inside the mushroom  (Figure \ref{fig:panels}) . Fused distribution blocks with custom cabling connects the power to the cameras. The filter wheels use a similar, but smaller  120V input / 12V 8A output power and distribution located on the same panels. An additional 120V input / 12V 8A output power supply is also available on each panel to supply the focus servos, cooling fans, and other accessories. A panel attached to the center of the base over the mount (Figure \ref{fig:panels}) holds a Network Power Supply (NPS) and a power supply for the USB hubs used to control the cameras assemblies. The selection and placement of the power systems allows for proper balancing of the mushroom assembly, cooling of the electronics, access to all of the components, and provides a safe supply of power to many different systems confined in a small area.
   
\subsubsection{Cooling}
The Evryscope uses up to 1.2kW when all cameras are cooling at maximum power, producing a significant amount of heat within a 6-ft semi-enclosed space. In-lab tests showed that parasitic heating between cameras could lead to a thermal runaway under some environmental conditions: cameras pulling in warm air exhausted by the thermoelectric coolers of neighbouring cameras must work harder to cool their sensors, increasing the amount of waste heat exhausted, and causing other cameras to further increase their cooling power. This process headed for runaway when the air temperature inside the mushroom exceeded $\approx32^{\circ}C$. Although several layers of protection prevent hardware damage from overheating (\S~\ref{rcs}) this could have impacted system uptime during summers. 

We implemented three systems to eliminate the parasitic heating. First, we built aluminum deflectors to move the camera exhaust air towards the center of the mushroom. Second, we added a bank of 8 120mm low-vibration 12V fans to direct cool air to the top of the mushroom. Third, we added external Vornado high-volume industrial fans to direct large amounts of external cool air to the mushroom (when rarely necessary). Together, these systems produce a coherent flow of cool air from the front-bottom of the mushroom to the top of the dome and down again out of the back of the systems. Testing showed no measurable effect on image quality when all systems are activated. The thermal protection systems have not triggered a shutdown since this system was commissioned. 

\subsubsection{Environmental Monitoring}
We monitor the hardware status with sensors distributed around the mushroom and dome, all linked to the main control system via ethernet or USB connections. The main control computer runs automated analysis and control scripts, and alters the state of fans as necessary to maintain stable temperatures around the cameras. Logs of all sensor values are recorded each minute.

Inside the mushroom, each camera has an external temperature sensor, measuring the air temperatures at 22 points around the dome. An environment-monitoring Raspberry-Pi is located at the center of the mushroom. Its custom-built sensor board monitors the overall mushroom temperature with a wide-angle infrared thermometer, the center-mushroom temperature with a built-in sensor, and the tilt of the mushroom using a precision three-axis accelerometer. A timing GPS system is also connected at that location. A summary of all sensors is shown in Table \ref{tab:sensors_tab}.

\begin{table}[htbp]
\caption{The Evryscope Environmental Monitoring Sensors}
\centering
\begin{tabular}{ll}
\hline
Description & Location\\
\hline
Mushroom interior temperature & 22 sensors in cameras\\
Overall mushroom temperature & Watchdog RasPi\\
Mushroom electronics temperature & Watchdog RasPi\\
Three-axis-accelerometer tilt & Watchdog RasPi\\
GPS timing sensor & Watchdog RasPi\\
Webcam dome light level sensor & Dome control RasPi\\
Rain sensor & Dome control RasPi\\
Smoke detector & Dome floor\\
Pier-base temperature sensor & Mount controller\\
Weather station & PROMPT array\\
\hline
\end{tabular}
\label{tab:sensors_tab}
\end{table}

Outside the mushroom, two webcams continuously monitor the system from the North and the South. The Northern webcam is a pan/tilt unit; the Southern webcam is a Raspberry-Pi camera which, in addition to providing a view of the mushroom internals, automatically monitors the light level in the dome. If the light level is consistent with the dome being unexpectedly open in daytime, a loud alarm bell is sounded and the Evryscope team is alerted via email. 

We use the PROMPT weather monitoring system \citep{2005NCimC..28..767R} for dome open/close decisions; this system has been in reliable operation for almost a decade. The PROMPT weather station monitors cloud levels, wind, and dewpoint. We use the RASICAM \citep{2010SPIE....10.1117} system to log cloud measurements for data-quality testing.

\subsubsection{Data \& control signal distribution}
The main control computer, watchdog and environment-monitoring computers and data-storage and analysis servers are located within the telescope dome, with optical fiber connections to a backup storage site in an adjacent PROMPT dome. The Evryscope data and control bus is a gigabit ethernet system operating as a separate subnet behind a router connected to the main CTIO network. 

A single sealed and fanless Logic Supply ML600G-30 rugged computer runs the robotic control software (\S~\ref{rcs}) and the USB-controlled devices, including the cameras, filter wheels, Robotilters, and the mount.  

Over 50 individual USB devices are connected to the control computer, which produces challenges to reliable system operations (ethernet control was not available for our chosen cameras at the time of system design). We initially connected groups of 4-8 USB devices together using powered USB hubs. However, lab testing showed occasional USB-bus-voltage brownouts, where the 5V power supply in a typical computer could be pulled out of voltage specification just by connecting dozens of USB devices, even when the devices were powered off and connected via powered hubs. This could prevent the control computer starting up or cause unreliable operation, and occurred for all tested brands of USB hubs. We eliminated this problem by finding and removing an undocumented jumper inside Starlink ST7200USBM rugged USB hubs which completely disconnects the upstream USB power rails from the downstream devices; this produces reliable operation with at least 60 USB devices connected.

\subsection{Robotic control software} \label{rcs}
The Evryscope is controlled by custom Python framework running on several computers within and outside the mushroom. We use a daemon-based software model, where each subsystem is controlled by an individual script operating as a separate process; this ensures that crashes related to individual hardware components do not stop the control of the other components. Critical systems such as emergency watchdogs are located on separate computers, allowing the entire system to enter a safe mode in an emergency even if the main control computer is disabled. The 18 daemons comprise 18,000 lines of Python code and communicate via a JSON-based protocol on TCP/IP sockets. 

\begin{figure}[tbp]
\includegraphics[angle=-90, width=1.0\columnwidth]{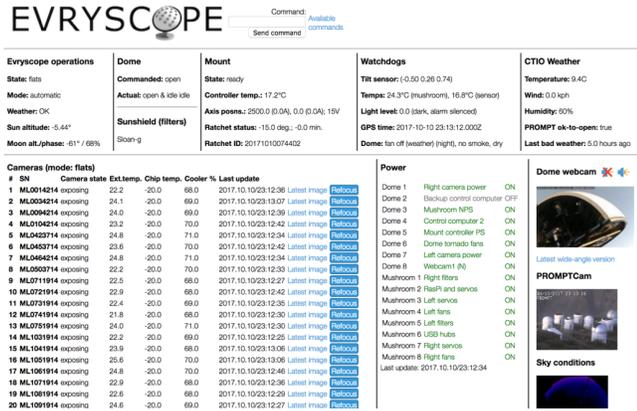}
\caption{The Evryscope status webpage, used for system monitoring and control. Commands can be issued to each hardware and software system using buttons or a simple text interface.}
\label{fig:status_page}
\end{figure}

A supervisory daemon is responsible for overall control, working as a finite-state machine to decide on the current best system operation mode from a range of options (science operations, taking calibrations, waiting for good weather, waiting for sunset, resetting mount for the next ratchet, and emergency shutdown mode). Transitions between modes are handled automatically by issuing commands to the relevant daemons and waiting for confirmation of hardware states as necessary. Commands to the hardware daemons range from simple (changing a filter position for example), to complex operations that can take many hours and involve large amounts of computing resources (executing a 3D-surface focus map for a camera, for example). A manual mode allows humans to issue commands directly to each daemon as necessary using the Evryscope status webpage (Figure \ref{fig:status_page}), although the supervisory daemon must be informed, or the unexpected hardware states will be detected as error conditions.

The system is designed to fail-safe, entering a safe mode on all important errors. Each subsystem daemon is responsible for the safety of its individual hardware components. This is relatively trivial in the case of filter wheels and similar low-impact systems, but is safety-critical for some components like the dome, the camera power supplies and the tracking mount. To produce a fail-safe mode, where the hardware is protected in the case of a system error or unexpected condition, the supervisory daemon issues a "heartbeat" ping to each daemon every 15 seconds. If the ping is not received on schedule, each individual daemon enters a safe mode -- closing the dome, powering off the cameras, placing the filter wheels into sunshield position (\S~\ref{subsub_sunshield}), and so on. Conversely, if a daemon does not respond to the heartbeat ping, suggesting it has crashed, the supervisory daemon triggers an error condition and stops issuing heartbeats to the other daemons. On any unhandled error condition the entire system enters semi-safe mode within $\approx$ 15s (dome closed, mount stopped), and fully-safe (sunblocks enabled, cameras powered off) within a minute. When this occurs, an email is sent to the Evryscope team for manual checks. This typically occurs once every few months, usually because of a communications glitch with an external component.

\subsection{System robustness \& failure mode mitigation} \label{system_robust}
The Evryscope is designed for fully-robotic operation with minimal on-site support. A rigorous analysis and mitigation of potential failure modes is vital to ensure robust operation. We categorized possible failure modes into a) problems that would allow the system to keep running with degraded performance and b) catastrophic failures that could cause permanent hardware damage. For the first type, we designed the system control software to monitor all hardware systems continuously and fail-safe into a known-good state on detection of errors (see \S~\ref{rcs}). For the potentially catastrophic problems, we designed multiply-redundant backup systems:

\subsubsection{Fire} The Evryscope uses up to 1.2kW of power when all systems are simultaneously operating, within a fairly-small enclosure. Two 120A/12V power supplies supply power to the camera systems, and a short-circuit on a 120A-capable line could easily produce enough heat to ignite surrounding material. We mitigated these concerns by a) breaking apart the high-current lines very close to the power supplies for individual camera power; b) individually fusing each power supply line; c) powering the system via GFCI breakers to produce a rapid shutdown in the event of a ground fault; d) specifying all plastics to be flame retardant; e) wrapping all exposed cables in flame-retardant material; f) placing an omni-directional infrared temperature sensor in the dome which shuts the power down on detection of an overheat condition; g) placing a Raspberry-Pi connected smoke detector in the dome to rapidly shut off power and sound an alarm if smoke is detected.

\subsubsection{Lightning} Electrical storms are rare at CTIO, but the Evryscope has so far experienced one extremely-nearby lightning strike that damaged equipment in nearby domes. To mitigate the possible lightning impact, we applied surge protectors to every power line and isolators to every USB and ethernet cable longer than three feet; this also mitigates the effects of possible ground loops. No lightning damage has been experienced by the system.

\subsubsection{Earthquakes} Chile regularly experiences large earthquakes, and telescope systems must be designed to survive large ground accelerations. As with the other main instrument components (\S~\ref{telescope_structure}) we evaluated the Evryscope pier mount design using 3D modeling finite element analysis. We simulated the telescope weight on the pier design over several angles to mimic positions during the ratchet cycle. The final pier design is 1/2" wall structural grade steel box tubing, with a strength failure several orders of magnitude above any level the Evryscope is likely to see. An accelerometer inside the mushroom measures the tilt of the mushroom and any other accelerations, and places the system in safe mode if limits are exceeded. On September 16, 2015 CTIO was hit by a magnitude-8.3 earthquake at a distance of 115 miles. The Evryscope automatically went into safe mode; no structural damage occurred and after quick manual checks the system was able to restart with no maintenance required.

\subsubsection{Sun exposure} \label{subsub_sunshield} With a telescope pointing at almost the entire sky, if the dome is opened during the day at least one camera would be pointing directly at the sun. The resulting heat buildup in the sun-pointing region of the CCD chip would be likely sufficient to cause significant CCD damage. If the dome was left open for an entire day, during maintenance or as a result of equipment failure, it is possible that an entire row of cameras could be damaged or destroyed. We addressed this with 1) a daylight alarm which sounds a loud bell and contacts the Evryscope team; 2) sunshields built into each camera. 

The sunshields are contained within the cameras' filter wheels and consist of a 3mm-thick steel washer backed by a mirror; sunlight entering the lens will be very out of focus at the filter position, preventing the formation of hotspots. Experiment at Chapel Hill showed no dangerous heating of the lens over hours of sun exposure. The sunshields are a primary safety system and as such are engaged immediately upon error conditions; each morning the system engages the sunshields as part of the shutdown procedure (apart from fans, the sunshields are the only moving parts inside the mushroom that are used nightly).

\subsection{Data analysis}
Here we describe briefly the Evryscope data analysis pipeline, forced-aperture photometry, and light curve generation; a full description will be published in upcoming work (Corbett et al. in prep.). As with many wide-field surveys, the Evryscope data analysis platform adapts established methods into a custom solution. The extremely wide field, concomitant optical distortions and flat-fielding challenges, and the very large quantity of data are the primary challenges. Each night, the Evryscope opens up and takes calibrations and science images automatically.  15-20 darks and twilight flats  are taken each night for each camera and on a typical observing night, with good weather, each of the 22 cameras will take 250-300 science images.

\subsubsection{On-site data analysis infrastructure}
The Evryscope generates approximately 6500 55MB science images each night. This data volume precludes transmitting the data for off-site processing with the current CTIO internet link. All data is therefore stored and processed on-site. Images are stored in an FPACK-compressed format across multiple Synology DS-2415+ network storage appliances, each of which is equipped with twelve 8 or 12 TB drives. In addition to image storage, we have provisioned a separate data store exclusively for our photometry database, consisting of 12 helium-filled 8 TB drives directly attached via a SAS backplane to our database server.

Data processing is split between two servers, both housed in the PROMPT domes at CTIO. The original server, a 12-core Intel Xeon based machine, was installed with the system. Post deployment, the mainboard of this server suffered some mechanical damage, limiting its RAM capacity to 112 GiB. In January of 2016, a second server was installed and the original was reprovisioned to support a calibrations and image indexing database, while all other analysis tasks were migrated. The second server is also based on the Intel Xeon platform, with 36 physical cores and 256 GiB of RAM. 

\subsubsection{Pipeline design}
The Evryscope currently runs a forced-aperture-photometry pipeline. The pipeline takes incoming images, calibrates them with darks and flats, generates a precision astrometric solution from the bright stars, estimates local background light and noise across each image, and measures aperture photometry for all sources from a reference catalog.  The Evryscope pipeline consists of $\sim$ 50,000 lines of custom Python and C++ code, with custom code performing flat-fielding, astrometric distortion correction, local background and noise estimation, precision aperture photometry, transient detection, and large-volume data storage. We expect to upgrade the pipeline to full image subtraction in the future.

We extensively tested standard data analysis software with Evryscope images (for example, the SExtractor \citep{1996A&AS..117..393B} and astrometry.net \citep{2010AJ....139.1782L} software suite used in PTF \citep{2009PASP..121.1395L} and the AWCams \citep{2013AJ....145...58L}). However, we found that the standard software struggles with our crowded images with large lens distortions: astrometry.net had a $>20\%$ probability of failing to find a good astrometric solution at the edges of the frames, often producing distortion solutions several pixels off. SExtractor often could not attain a good background noise estimate for our crowded images, and therefore set the source-detection requirements extremely high; often several-degree-wide regions of the Evryscope images did not show any detections despite tens of thousands of stars being clearly visible by eye. A few percent of the Evryscope images also showed SExtractor photometry very divergent from adjacent images, with stars' brightness measurements changing by tens of percent with no discernible by-eye difference in the input images; these problems persisted regardless of the input settings. For these reasons we developed a completely-custom pipeline, although we do use astrometry.net for initial rough astrometric solution and SExtractor for quick source-detection for camera focusing; both codes work very well for those applications.

Each processed night consists of $\approx$360GB of raw imaging data, resulting in several hundred new data points for each of $\approx$10M stars. On our current computing hardware, the pipeline is capable of processing $\sim$7 nights (2.5TB of imaging data) every 24 hours. This speed is necessary to allow us to re-reduce our current three-year dataset in a reasonable time. 

\subsubsection{Image quality checks \& calibrations} \label{image_quality_checks}
Each Evryscope science image is subjected to an initial quality control script which evaluates the image quality based on the presence of stars in the image, PSF shape (avoiding rare tracking errors), and background levels. Images that pass ($>$90\%) are masked for known bad pixels and columns.

Darks are taken daily with the filterwheel in the closed position, and monthly midnight darks are taken for comparison to check for light leaks. Masterdarks are generated by combining and median averaging several hundred darks. Our CCD characteristics are sufficiently stable to use the masterdarks for a season. 

Twilight and sunrise flats are taken daily and evaluated with a quality control check for stars and clouds. Residual point sources are removed. Lens vignetting and small scale interpixel variation in CCD sensitivity are removable to the one percent level with standard flattening procedures, however the large scale sky gradient due to the extremely wide field of view necessitates a more complex procedure. We constrain the large-scale variations on using on-sky photometric measurements of starfields, and measure the small-scale variations from the high-frequency structure in twilight flats.

\subsubsection{Photometry and light curves}
Our current dataset includes 9.3M stars with an average of 32,600 photometric measurement points. The photometric points are stored in a flat-file based custom backend storage system written in Python. The system is partitioned by sky position using HEALPix pixels \citep{Gorski_2005}. HEALPix pixels divide the sky into equal area regions; we selected a 3.5 sq. deg. HEALPix pixel size for convenience to limit the number of stars in a particular region. This aids in processing of the light curves (done per HEALPix pixel) and allows for multi-threading and tiling the database writing steps. We evaluated database management systems (DBMS), but found that for our extremely-consistent-format numerical data our custom system could reduce storage requirements by a factor of five compared to PostgreSQL while increasing access speed by a factor of ten. We also evaluated similar commercial and open-source flat-file numerical data storage systems  and found that the performance was generally comparable to our flat-file-based system, but with significantly higher implementation complexity and programming overhead. The flat-field storage system stores approximately 15TB/yr of light-curve data.

Each star's photometry is measured in five different photometric apertures, allowing an optimization of the SNR for each star (for example, selecting larger apertures for brighter stars; this technique is used by several surveys, e.g. \cite{2006PASP..118.1407P}). Each measured data point also includes the star's measured RA \& Declination, CCD position, estimated SNR, limiting magnitude at that point, background light level, peak flux level, and a GPS-based precision timing signal with tested 1~s accuracy \citep{2018AAS...23115023C}.

We periodically generate precision light curves for each star based on the typically tens of thousands of photometric points recorded for each star. The light-curve generation code processes each HEALPix pixel separately, performing differential photometry on the contained group of several thousand stars. Atmospheric extinction variations from clouds and airmass are corrected for using differential photometry among the thousands of stars in each HEALPix pixel. First, images pass through an image quality check which rejects images with high background, low numbers of detectable sources, or suspect PSF shapes. Next, the least-variable stars are automatically selected to form a consistent set of reference stars (this procedure is iterated with the differential photometry to find the stars most indicative of the overall photometric variations). For each single-camera image accepted by the pipeline for processing, which typically have a few 100,000 stars, each source is checked for possible blending, local background issues, non-detection and saturation. Flags are issued for suspect data points. Flux errors are estimated based on the local background noise for all epochs, for all sources. Airmass and differential chromaticity errors are removed by SysREM \citep{2005MNRAS.356.1466T} in the default pipeline operation; we tested removing explicit correlations with star color and measured airmass, but did not find a significant improvement in photometric precision. These procedures work for the large majority of the dataset, but a small fraction ($<20\%$) of the epochs are subject to largely un-removable variability due to thin clouds with spatial scales smaller than a HEALPix pixel. We detect and remove these epochs by searching for periods of higher-than-average photometric variability among all sources in the healpix, as well as higher-than-average extinction. We are currently developing methods to instead flag and recover these epochs for usable data.

We have implemented several layers of systematics removal, which can be applied depending on the science goals. All light curves are automatically decorrelated by two iterations of SysREM \citep{2005MNRAS.356.1466T}. Further iterations of SysREM further remove systematic errors, but there is also a risk of removing astrophysical variability. If only short-term variability is to be measured, such as in a transit or eclipse search, we add decorrellations of photometric variability with CCD chip position and airmass. We found that some long-term variables such as low-amplitude long-period rotation curves correlate with those telescope variables, and so we offer users the option of using uncleaned light curves. 

Processed light data is inserted to a PostgreSQL database, also partitioned into HEALPix pixels to increase performance. This database does not include much of the per-epoch metadata, and only contains results from the optimal photometric aperture. Each of the 6000 populated HEALPix pixels contains 0.2-2GB of light curve data, for a total light curve database size of $\sim$10TB. We query the database for target groups, and download the results to Chapel Hill for astrophysical analysis.


\section{Performance} \label{section_performance}

\subsection{Operations statistics}
The Evryscope saw first light on May 20, 2015 and has been operating continually since then with only brief maintenance shutdowns. From first light to August 1, 2018, 15.9\% of the nights were missed due to weather and equipment issues and 2.3\% of the nights were skipped due to planned maintenance. The maintenance trips occurred during November 11-20, 2015 (Robotilter installation and camera alignment); January 4-15, 2017 (lens cleaning, data storage increase, second analysis server installation, and general maintenance); and July 18-25, 2018 (lens cleaning and general maintenance). The fail-safe shutdowns occurred for the following reasons: excessive heat warning (20\%), dome control warnings (33\%), and smoke/dust/other warning (47\%). Almost all of the fail-safe shutdowns were false alarms, but we designed the system to be conservative with the goal of detecting real danger situations at the expense of some false positives.
	
\subsection{Hardware reliability}
The Evryscope has operated reliably for over three years, with only minor hardware issues. The mount has tracked over 5700 2-hour ratchet cycles with no major problems; during the 2017 maintenance trip we greased and tightened the worm gear adjustment which helped smooth the mount operation at peak stress positions. The support structures, including the fiberglass mushroom, have been durable and shown no signs of excessive wear or stress. The power supply units (cameras, filter-wheels, servos, USB hubs and accessories) have all performed without issue. The cameras have also run reliably and without failure. Three filter-wheels have failed over the course of three years. One broke a drive chain, while the other two stuck during routine cycling. One was stuck in the Sloan-g position so it did not affect imaging, the others were stuck closed so we lost the ability to image with two cameras until the next maintenance trip. One power cable to a camera USB hub failed in mid 2018 which disrupted operation of four cameras and filter-wheels; it was easily replaced during the June 2018 maintenance trip. The system is well sealed and minimal dirt and dust accumulates inside the mushroom. The optical windows need to be manually cleaned each trip, but the lenses can be cleaned simply with compressed air and/or off-the-shelf DSLR camera lens-cleaning pens.

\subsection{Imaging Performance}
The Evryscope imaging performance sets the limiting magnitude, photometric performance, and ease of source separation and image subtraction. In this subsection we explore the system's performance over the first three years of operation.

\subsubsection{Point Spread Functions}
\begin{figure}[tbp]
\includegraphics[width=1.0\columnwidth]{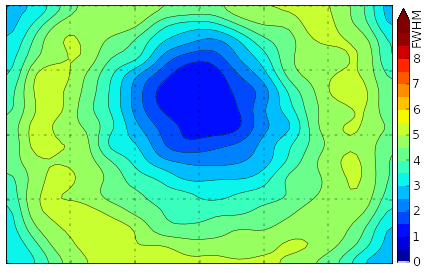}
\caption{The FWHM map of the camera pointing toward the South Celestial Pole. The image quality shows little tilt and a symmetric pattern.}
\label{fig:FWHM_map}
\end{figure}

\begin{figure}[tbp]
\includegraphics[width=1.0\columnwidth]{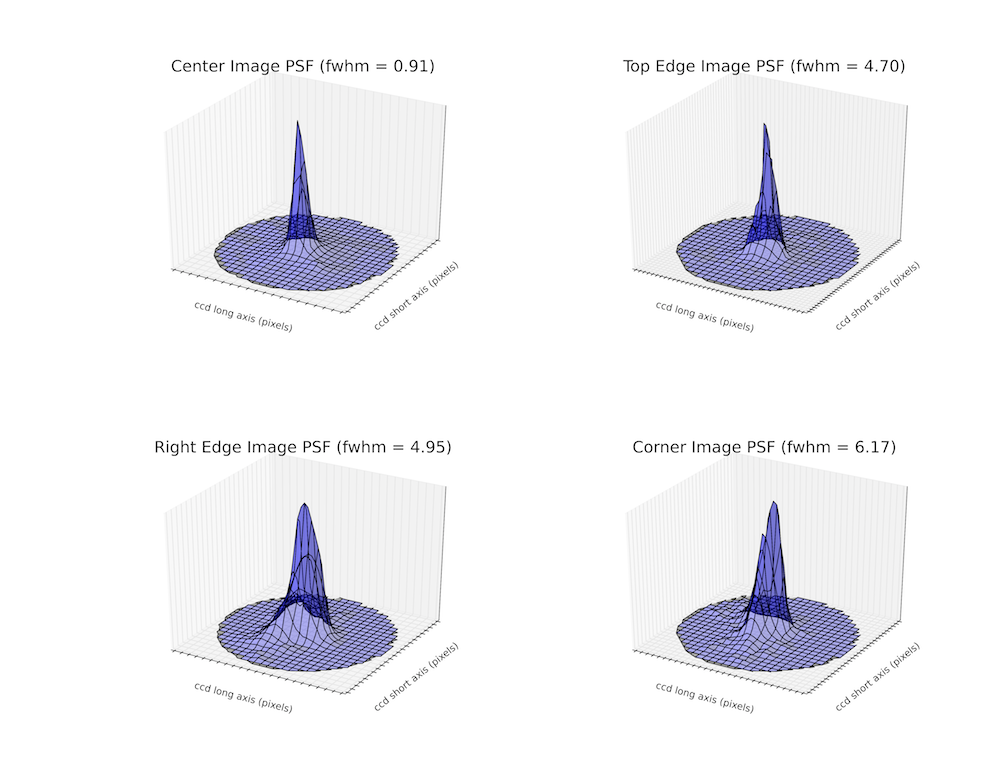}
\caption{Example medium brightness stars' PSFs from the center, edges, and corner of a representative camera.}
\label{fig:PSFs}
\end{figure}

The Robotilter camera/CCD automated alignment system is designed to remove tilt, minimize PSF distortions, optimize the focal plane, and defocus the image center. The FWHM (PSF full-width-at-half-maximum) map of a well aligned, representative camera is shown Figure \ref{fig:FWHM_map}).  Very little tilt across the image is evident, and PSF widening toward the corners due to lens coma, focus, and vignetting is within the expected range for our lenses. The PSFs range from 1-5 pixel FWHM  across much of the image -- 60 percent are less than 4 pixels and 90 percent are less than 6 pixels. Figure \ref{fig:PSFs} shows point spread functions for the central region and edges of a representative camera.

\subsubsection{Limiting magnitudes and coaddition}
\begin{figure}[tb]
\includegraphics[width=1.0\columnwidth]{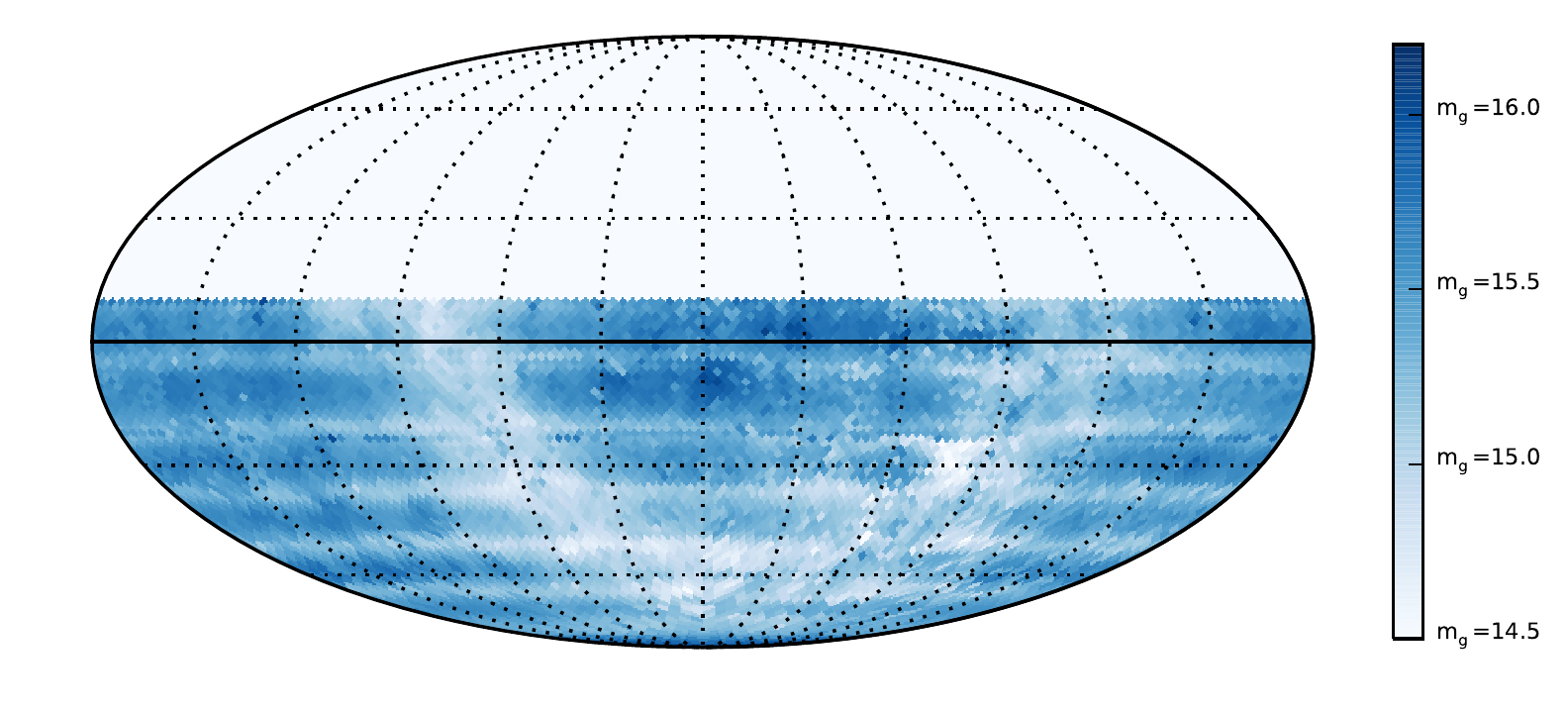}
\caption{The median dark-sky limiting magnitude for Evryscope data, measured in $\approx$32,000 epochs over three years of operations. The crowding effects of the galactic plane are visible, along with the striping from falloff in PSF quality towards the edges of the cameras' fields.}
\label{fig:limiting_mag}
\end{figure}

We calculate the limiting magnitude achieved by the system in each epoch by taking the faintest stars in each healpix and fitting the SNR decrease as a function of the g-band magnitude as measured by APASS. The dark-sky limiting magnitude (Figure \ref{fig:limiting_mag}) reaches our expectation of $m_{g'}\approx16$, with crowding from the galaxy reducing the limiting magnitude by approximately a magnitude in low-galactic-latitude areas. A horizontal stripe pattern is visible in the limiting-magnitude map; this is caused by the falloff in PSF quality towards the edge of camera fields of view.

\begin{figure}[tb]
\includegraphics[angle=-90,origin=c,width=1.0\columnwidth]{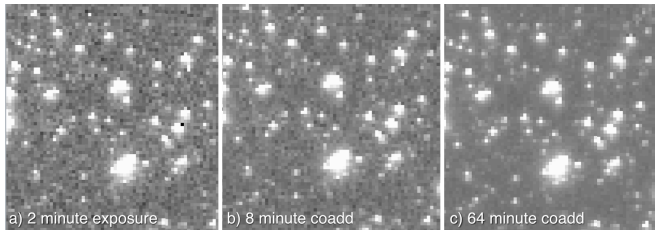}
\caption{Progressive coaddition of a selected sky region, with image scaling applied to show the noise structure in the images. As well as increasing depth, coaddition with the slow star position changes over a ratchet allows the removal of bad and hot pixels.}
\label{fig:coadd_depths}
\end{figure}

\begin{figure}[tb]
\includegraphics[angle=-90,origin=c,width=1.0\columnwidth]{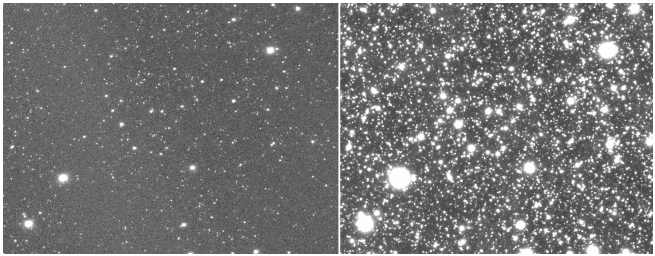}
\caption{\textit{Left:} a selected region of a single two-minute Evryscope exposure. \textit{Right:} co-addition of a full night of data from the same region, with scaling to show the increased number of stars and the bright-star PSFs.}
\label{fig:deep_coadd}
\end{figure}

The camera gains, data compression and calibration fidelities are selected so that coadding the data achieves greatly improved signal to noise, with depth increasing with approximately the square root of the number of exposures (Figure \ref{fig:coadd_depths}). In uncrowded regions of the sky during dark nights, the system typically achieves $m_{g'}=17$ in 8 minutes coadding (4 exposures), $m_{g'}=17.5$ in 32 minutes, $m_{g'}=17.8$ in 64 minutes, and $m_{g'}=18.5$ in 360 minutes (the latter crowding-limited over much of the sky; Figure \ref{fig:deep_coadd}).

\subsubsection{Photometric precision}
\begin{figure}[tb]
\includegraphics[width=1.0\columnwidth]{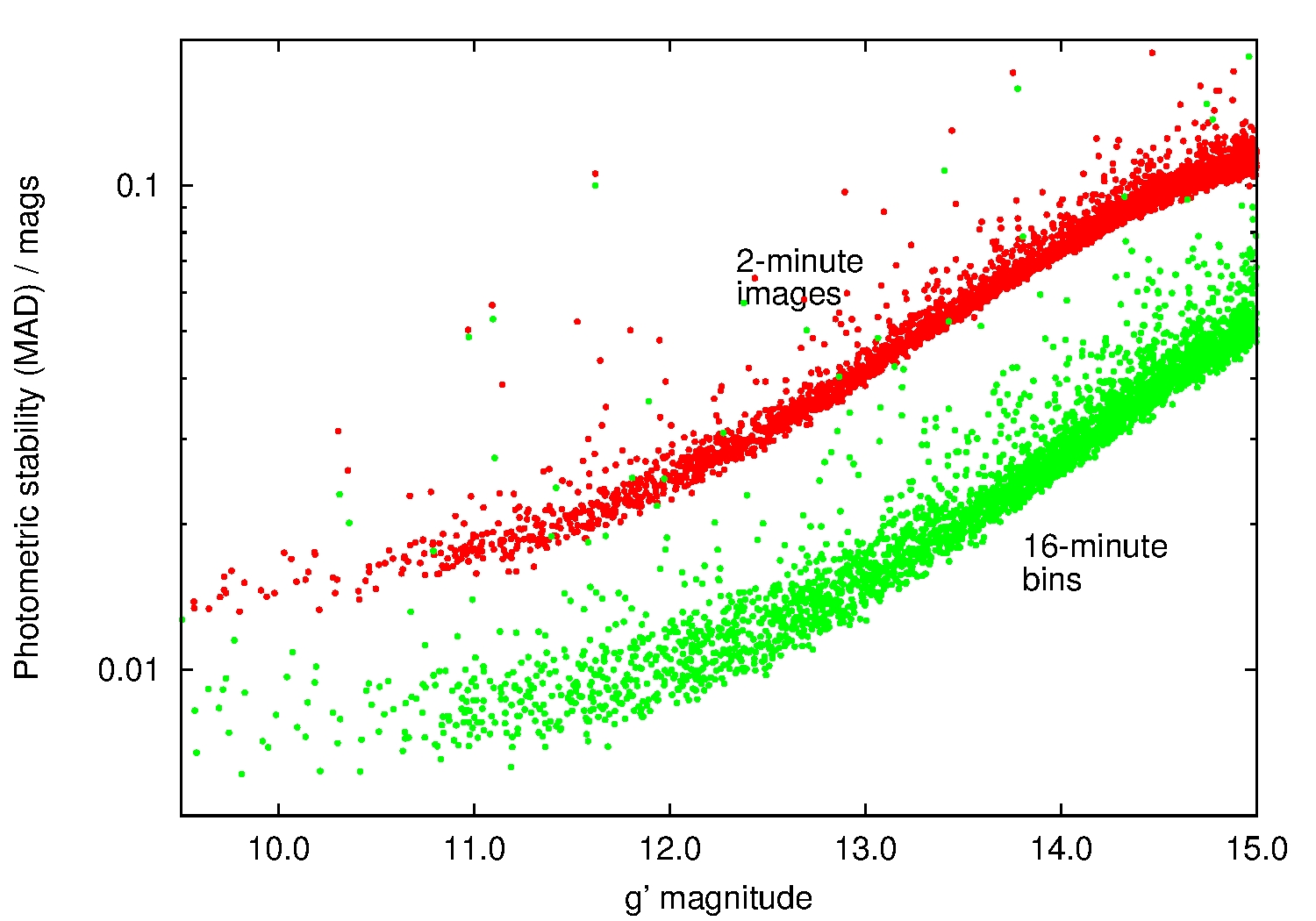}
\caption{Evryscope light curve photometric performance per magnitude for three years of data under all moon and cloud conditions. Stars in a  representative HEALPix pixel of the Evryscope database targets is shown for visual clarity. The high RMS outlier points are astrophysical variable stars.}
\label{fig:photometric_perf}
\end{figure}

Light curve performance reaches our expected performance levels of near 1\% rms on bright stars and $\sim$ 10\% on dim stars, over three years of data under all moon and cloud conditions (Figure \ref{fig:photometric_perf}). With binning and/or aggressive removal of poor conditions data and systematics, the performance is improved to the 6-millimag level. These levels are greatly improved when coadding epochs for the detection of periodic objects, where we have published clear signals at the few-millimag level \citep{2018AJ....156..120T}.
	
\section{Example light curves, discoveries, and on-going surveys} \label{section_results}
The Evryscope has a wide variety of on-going surveys (\S~2.1). In this section we detail results from some of the current surveys, provide example Evryscope light curves and discoveries from a selected region of the sky; many more comprehensive surveys are currently ongoing.

\subsection{Candidate Detection}
The Evryscope team uses a wide range of detection tools, given the variety in the science survey goals (see \S~2.1). Box Least Squares (BLS) \citep{Kovacs:2002gn}, \citep{2014A&A...561A.138O} is the primary search tool used for conventional (wide, shallow, many points) transit like detections. The box size, sampling, and period range are selected depending on the host star and expected companion type. To find potential transiting planets with compact host stars such as white dwarfs or hot subdwarfs, where the transit times are orders of magnitude shorter, we developed a custom code written in Python which we call the outlier detector. It excels in finding very short time (on the order of a few minutes to tens of minutes) transits with deep (ten percent or more) depths, even for faint objects. We use several iterative processes to select low outlying points and find the period with lowest in phase deviation. Flares are discovered and characterized with an automated flare-analysis pipeline which uses a custom flare-search algorithm, including injection tests to measure the flare recovery rate. The algorithm searches for flares by first dividing each lightcurve into segments of continuous observations and subsequently fitting an exponential-decay matched-filter to each contiguous segment of the light curve. Matches with a significance greater than 4.5$\sigma$ are verified by eye. Microlensing events are detected with a differential image / matched filter Python code that triggers an alert if required parameters are met. Lomb Scargle (LS) \citep{1975Ap&SS..39..447L}, \citep{1982Ap&SS..263..835S} is the primary algorithm used to find stellar variability and binaries.

\begin{figure*}[ht]
\epsscale{1}
\includegraphics[angle=90,origin=c,width=.60\textwidth]{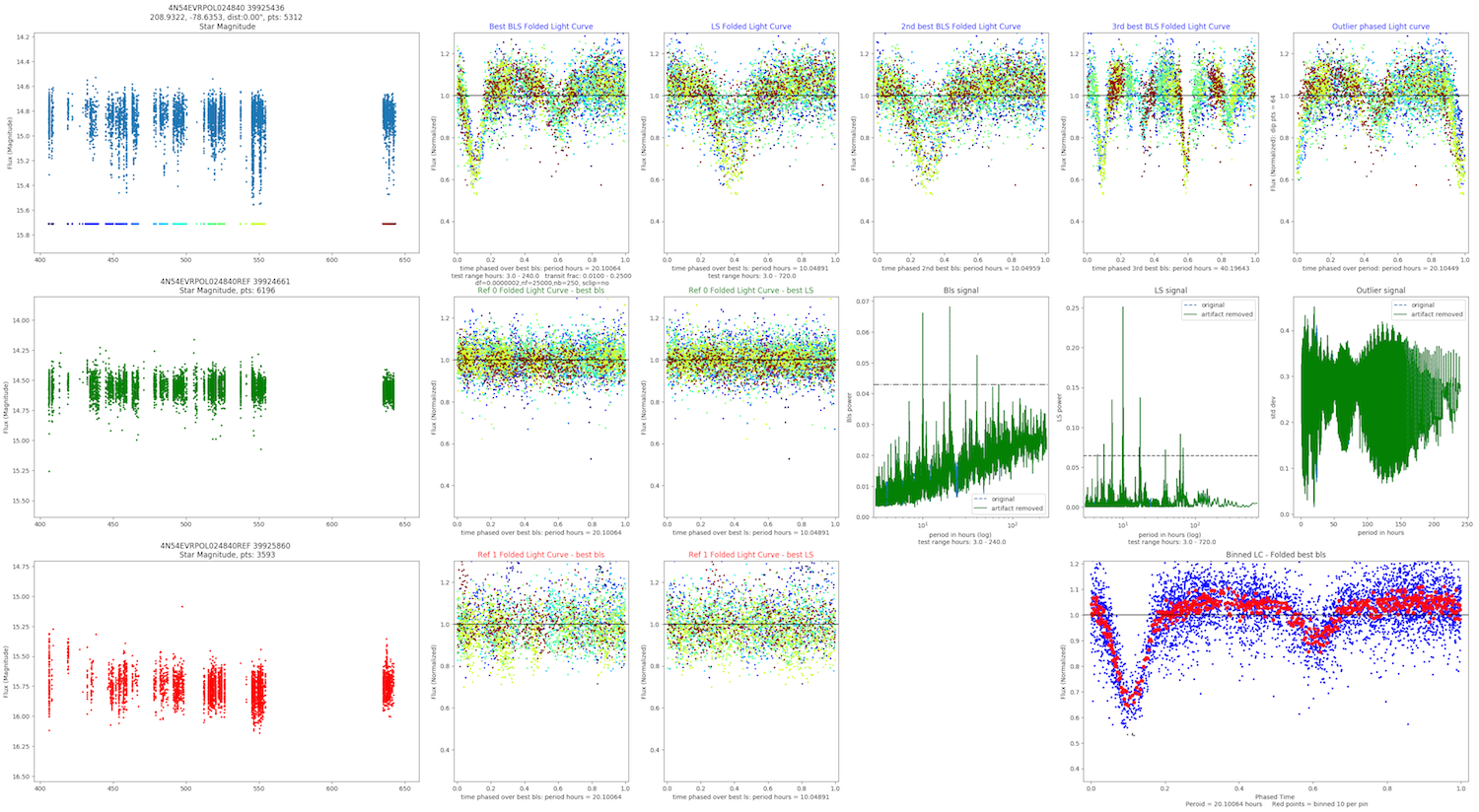}
\caption{Evryscope transit detection display panel, with a newly discovered eclipsing binary. The left panels show the target and two reference star light curves, as well as the BLS and LS phase folded on the best period. The coloring of points shows the mixing of the best period find and comparison to nearby references for identification of systematics. The right panels show the outlier results and the binned light curve folded on the best period.}
\label{fig:detection_panel}
\end{figure*}

Visual inspection and systematic assessment is a key to detection and false positive elimination. We have developed several visual tools including the display panel plot (Figure \ref{fig:detection_panel}) that allows for simple and effective visual confirmation of candidates. In the same panel plot, we test the candidates for signs of systematics by comparisons to nearby reference stars, examining binned data, and checking for alias and data gaps. Fit power, ordering, and selection of top targets is available to narrow the candidates depending on the search and number of targets. This display is available for all Evryscope light curves, on request.

\subsection{First discoveries}
The Evryscope team (and collaborators from 17 institutions) are engaged in a wide variety of astrophysical projects with the light-curve dataset. The first major Evryscope result, the first detection of a superflare from Proxima Centauri, was recently published in ApJL \citep{2041-8205-860-2-L30}. Several other papers are currently under review, and many more results in prep. Here we show some examples of variability discoveries from the Evryscope database, and results from a test search in a selected region of the sky. We follow with updates on the various surveys that are underway.

\subsubsection{New Eclipsing Binary / Variable Star Discoveries} \label{subsection_VSD}
\begin{figure*}[!h]
\epsscale{1}
\includegraphics[width=1.0\columnwidth]{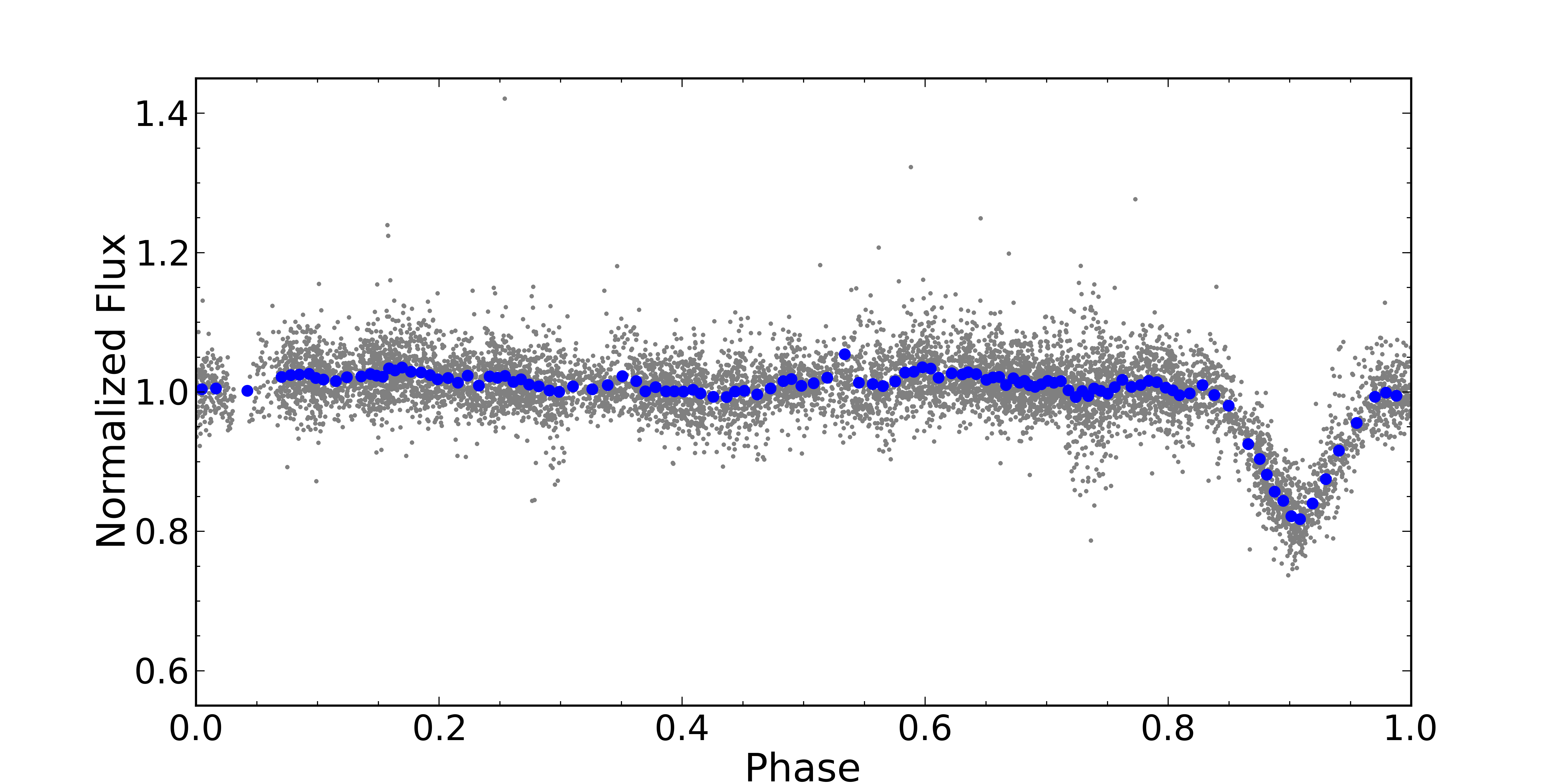}
\includegraphics[width=1.0\columnwidth]{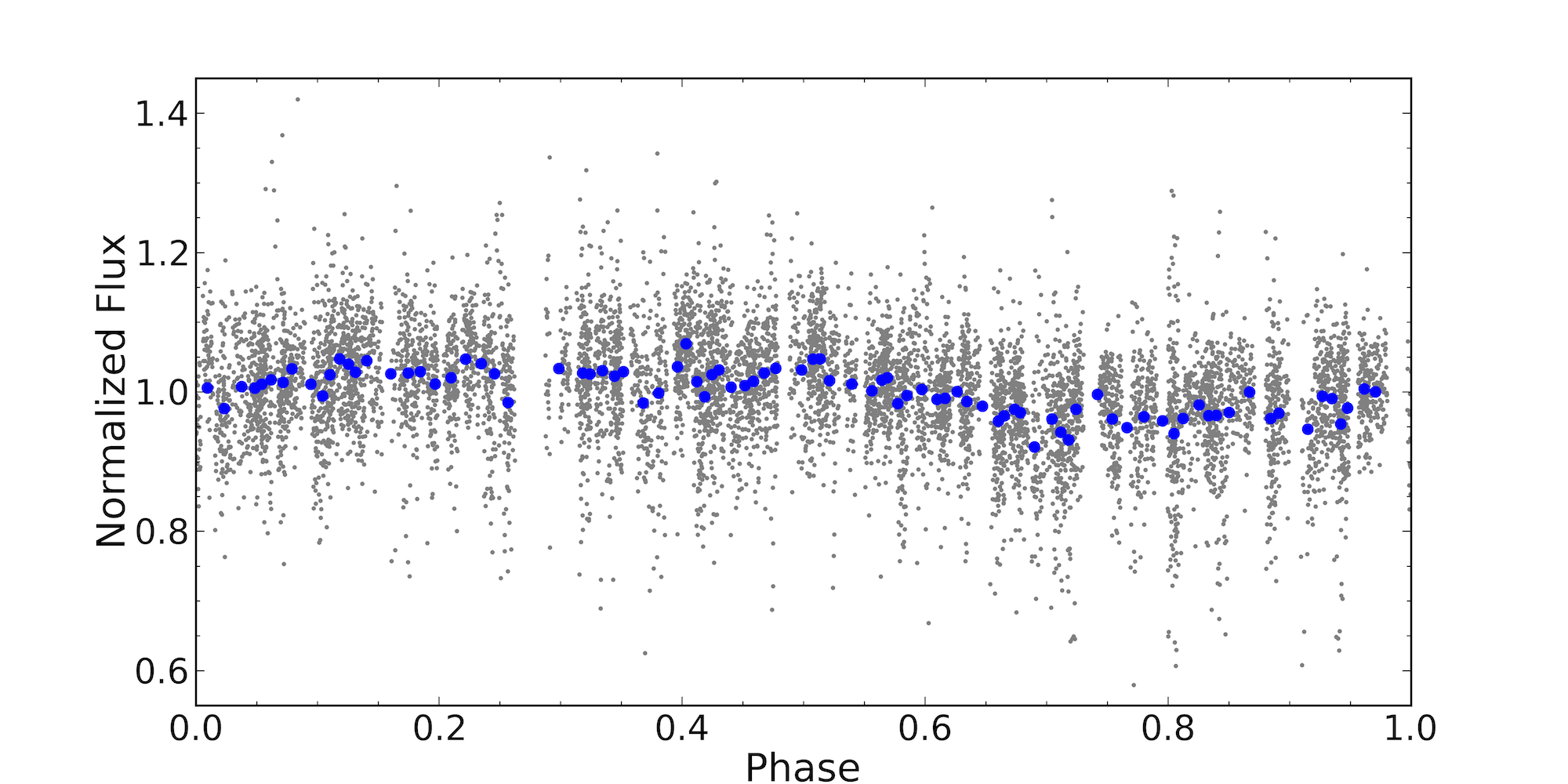}
\caption{Left: An eclipsing binary discovery folded on its 61.4905 hour period representative of 100's of Evryscope variable discoveries. Right: A variable star discovery folded on its 219.8386 hour period representative of 100's of Evryscope variable discoveries.}
\label{fig:example_1}
\end{figure*}

\begin{figure*}[!h]
\epsscale{1}
\includegraphics[width=1.0\columnwidth]{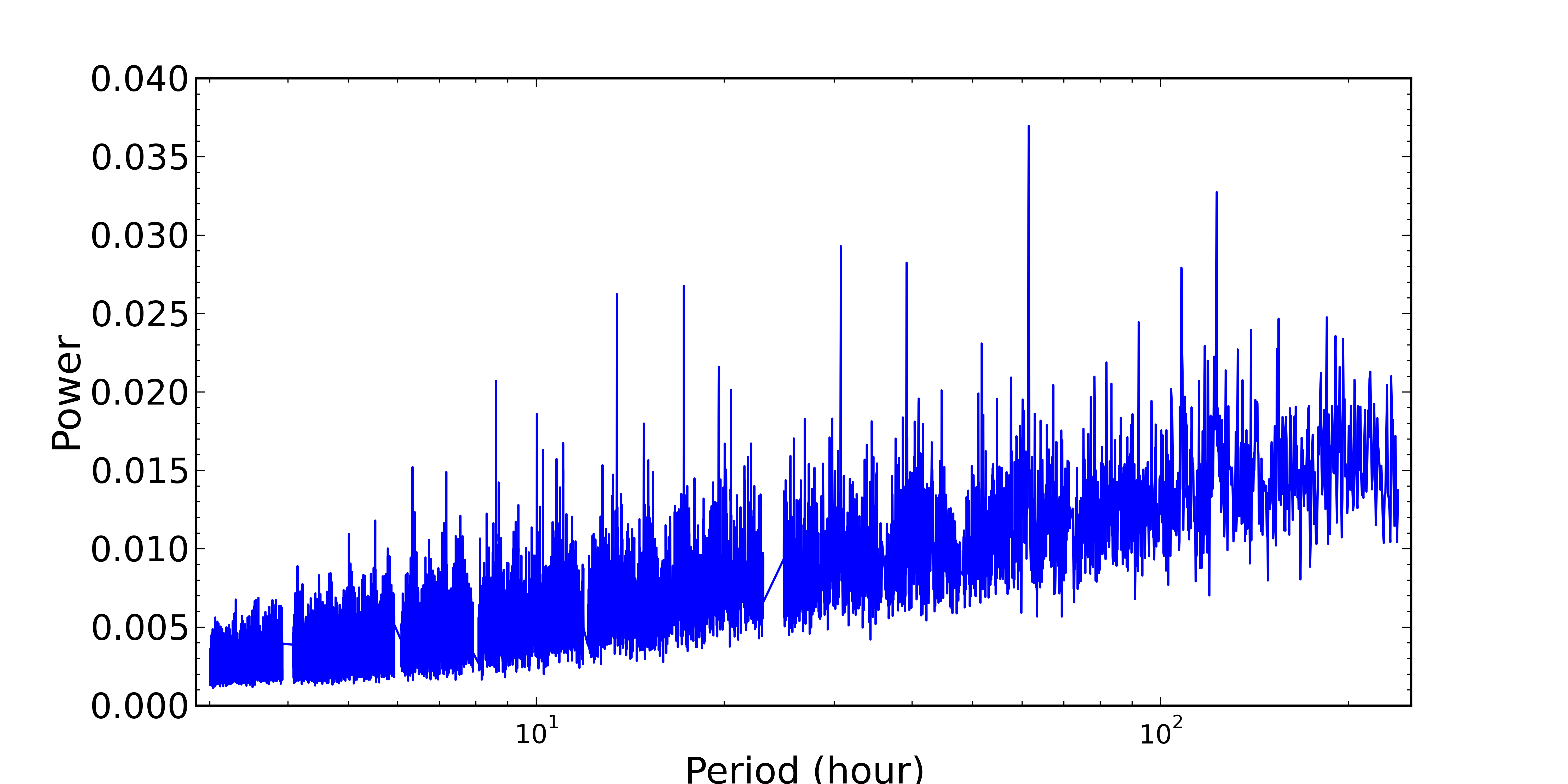}
\includegraphics[width=1.0\columnwidth]{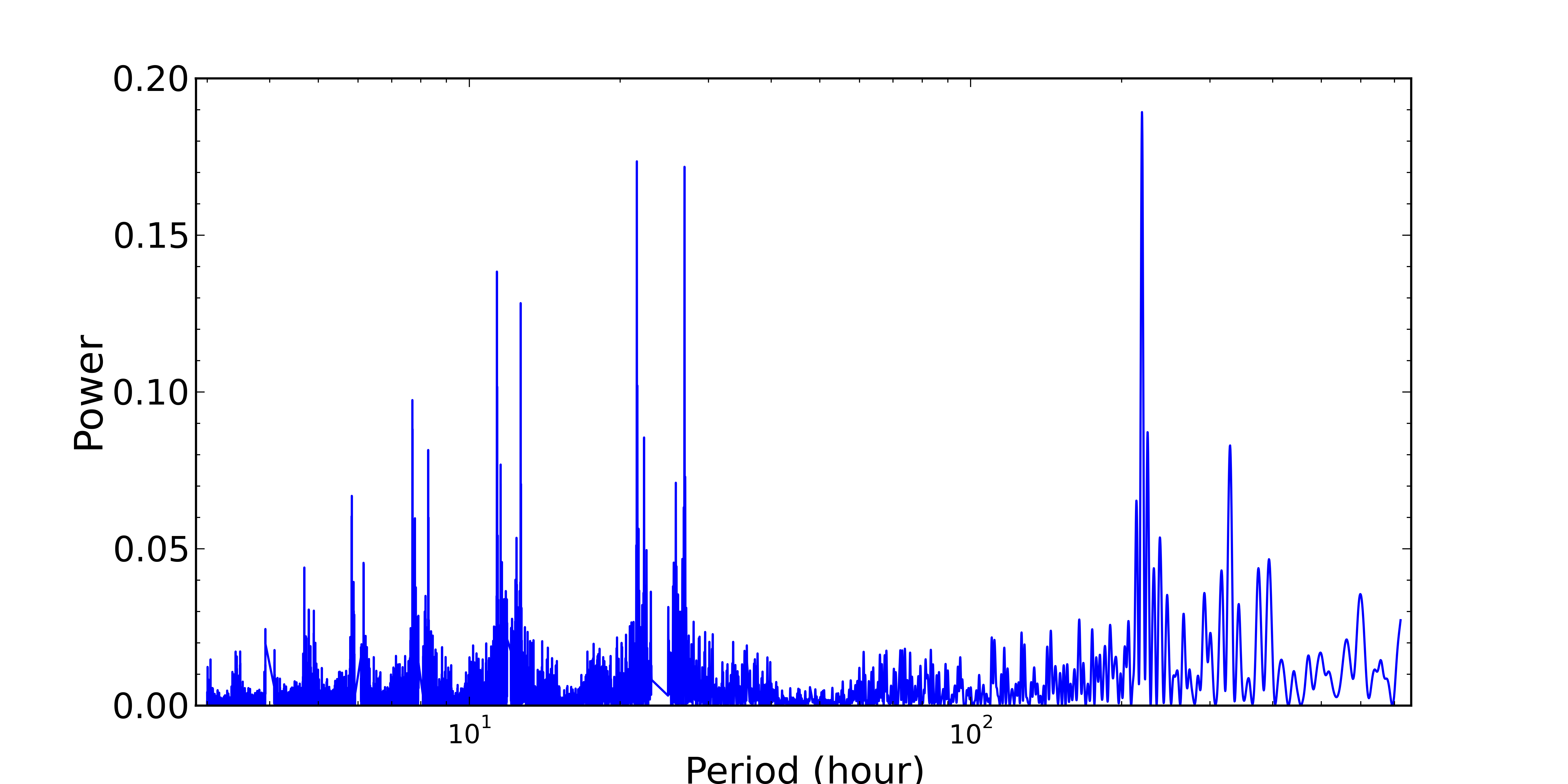}
\caption{Left: The BLS power spectrum (to the 61.4905 hour eclipse in Figure \ref{fig:example_1}) with the highest peak at the 61.4905 hour detection. Right: The LS power spectrum (to the 219.8386 hour variable star in Figure \ref{fig:example_1}) with the highest peak at the 219.5521 hour detection.}
\label{fig:example_1_power}
\end{figure*}

A test search limited to the northern region (declinations from +5 to +10), filtering the targets by magnitude (bright stars) and color (likely K-dwarfs or M-dwarfs) yielded 59 new eclipsing binaries and variables. Representative examples of an eclipsing binary and a low amplitude variable are shown in Figure \ref{fig:example_1}. The search was run by selecting all of the sources in the Evryscope database with light curves with greater than 5000 epochs, with magnitudes brighter than 14.5, and with sources that matched to PPMXL \citep{2010AJ....139.2440R} and APASS-DR9 \citep{2015AAS...22533616H} catalogs which could be classified as potential K-dwarf or M-dwarfs based on reduced proper motion (RPM) and B-V colors. After removing known variables, BLS and LS were run on the filtered list; the example eclipsing binary and low amplitude variable BLS and LS detections are shown in Figure \ref{fig:example_1_power}. The BLS and LS results were ordered by significance and the top 10\% were inspected using the detection panel plots. Those passing the visual inspection and systematics test were sent to the next stage. Eclipsing Binaries were fit with a Gaussian to measure the eclipse depth using the detected period and phase as the prior. Variables were fit using Lomb-Scargle to determine the amplitude. Example eclipsing binary and low amplitude variable fits are shown in Figure \ref{fig:example_1_fit}. Tables 4-6 in the appendix contain the full discovery list; Figures 23-24 display the light curves.

\subsubsection{Transit Surveys}

One major Evryscope transit survey has been completed and two are underway, with several others in the planning stages. A transit search for variable stars in the southern polar region led to 300 variable and eclipsing binary discoveries, with six of the eclipsing binaries having low-mass secondaries (Ratzloff et al., submitted). An exoplanet survey of $\approx$ 2500 southern sky white dwarf (WD) targets $m_{v} <$ 15.0 is underway. A transit survey of $\approx$ 3500 hot subdwarf (HSD) targets is in progress and has already discovered several rare systems: 2 HSD / low-mass-secondary eclipsing binaries (HW~Vir systems), 4 HSD reflection binaries, and 2 HSD / WD short-period binaries (all Ratzloff et al., in prep). From these surveys, there have been 5 planet candidate detections; subsequent followup showed these candidates to be grazing eclipsing binaries with almost identical stars or low-mass stellar companions. These detections demonstrate the Evryscope is capable of detecting planets orbiting post main-sequence stars as well as M and K-dwarfs with our current light curves and search algorithms. We have used the initial results of these first surveys to refine our transit searches; we briefly describe the status of the key Evryscope transit surveys below.

\textbf{White Dwarfs (WD):} Recent discoveries of WD debris discs and disintegrating planetesimals have fueled the speculation that planets could be present in WD systems \citep{2013MNRAS.432L..11L}, \citep{2018MNRAS.473.2871V}. WD exoplanets would have very short (few minutes to tens of minutes) transit duration and very deep ($\sim$ 100 percent for earth size planets) transit depths. WDs are extremely numerous in the sky as $>$ 90 percent of main sequence stars will eventually become WDs, however the low luminosity and small size make these stars observationally challenging. We leverage the Evryscope fast cadence and all-sky coverage to search for WD planets. Our first results from  $\approx$ 2500 southern sky WD targets $m_{v} <$ 15.0 did not return any candidates. We have improved our systematics removal, increased our coverage to 3.5 years, and added targets down to $m_{v} <$ 16.0 and will search again once the database processing is complete (Ratzloff et al., in prep). In the event of a null detection, we can provide upper limit constraints on WD planetary populations.

\textbf{Hot Subdwarfs (HSD):} HSD planet or low-mass-secondary transit durations are on the order of tens of minutes, and reasonably deep transit depths ($\sim$ 10 percent for Neptune size planets). A transit survey of HSD planets and other variability from a target list \citep{2017OAst...26..164G} of $\approx$ 3500 known HSD is in progress (Ratzloff et al., in prep). Although the survey is currently underway, several candidates, including the 8 mentioned above, have been identified and are pending further followup.

\textbf{M and K-dwarfs:} The Evryscope is capable of detecting $\sim$ 2 Earth radii M-dwarf planets and gas giant K-dwarf planets. A transit search for variable stars in the southern polar region detected a 1.7 $R_J$ planet candidate with a late K-dwarf primary. This system was later shown to be a grazing eclipsing binary, but demonstrated the Evryscope detection capability. An exoplanet survey of M and K-dwarf stars based on identifying candidates in our fields from spectral classification is planned for the entire sky when the HSD and WD surveys are completed.

\subsection{Other Variability Searches}

\subsubsection{Solar Flares and CME}

Flares and coronal mass ejections (CMEs) are capable of severely affecting the survivability of potentially habitable worlds. A comprehensive flare survey of M-dwarf stars (including known exoplanet hosts) of the southern sky is underway (Howard et al., in prep). These results, when combined with CME observations will be used to estimate the effects on long-term habitability of rocky planets orbiting M-dwarf stars. 

\subsubsection{Transient Detection}

We have developed tools for rapidly generating small cutouts from full-frame Evryscope images and performing high-precision photometry on uncataloged sources not included in our primary forced-photometry reduction, including difference image analysis for objects in crowded regions of the sky. This tool chain is designed to provide early pre-discovery photometry to help constrain the evolution of novae and supernovae. 

An example of Evryscope transient capability is a recent classical nova (Nova Carinae 2018) with pre-discovery Evryscope coverage \citep{2018ATel11467....1C}, which is currently under analysis and a detailed light curve will be presented in an upcoming paper (Corbett, et. al., in prep). The Evryscope data complements the later discovery by the All Sky Automated Survey for SuperNovae (ASASSN \cite{2018ATel11454....1S}) and the serendipitous space-based photometry of the Bright Target Explorer (BRITE \cite{2018ATel11508....1K}). High-cadence and high-coverage observations of classical novae can provide insight into the shock physics that drive light curve evolution \citep{2017NatAs...1..697L}. Also shown in Figure \ref{fig:example_transient_variable} is transient discovery from the variable star test search (\S~\ref{subsection_VSD}).

\section{SUMMARY} \label{section_summary}
The Evryscope was deployed to CTIO in May 2015 and has recently been joined by a Northern-hemisphere telescope at MLO. The Evryscope is designed to detect short timescale events across extremely large sky areas simultaneously. The 780 MPix 22-camera array has an 8150 sq. deg. field of view, 2 minute cadence, and the ability to detect objects down to $m_{g'}\simeq$16 in each dark-sky exposure. We have collected over 250TB of images and produced 25TB of light curves. In this paper we described the Evryscope hardware and explained why we designed the telescope as we did. The time from conceptual design to deployment was one year and the total hardware cost was $\approx$\$300K, meeting our time and budgetary goals. We demonstrated the on sky performance met our goals for telescope operation and reliability, sky tracking, threat mitigation, and reliability. Image quality reached our predictions for signal, noise, background, and PSF quality. The photometric pipeline produces light curves with the precision necessary to support the planned Evryscope science. We demonstrated the photometric performance by presenting select variable star discoveries and discussing rare hot subdwarf and white dwarf eclipsing binary discoveries. Updates on the status of our transit surveys, M-dwarf flare survey, and transient detection were also given.

This research was supported by the NSF CAREER grant AST-1555175, NSF/ATI grant AST-1407589, and the Research Corporation Scialog grants 23782 and 23822. HC is supported by the NSF GRF grant DGE-1144081. OF and DdS acknowledge support by the Spanish Ministerio de Econom\'ia y Competitividad (MINECO/FEDER, UE) under grants AYA2013-47447-C3-1-P, AYA2016-76012-C3-1-P, MDM-2014-0369 of ICCUB (Unidad de Excelencia 'Mar\'ia de Maeztu').

\begin{figure*}[!h]
\epsscale{1}
\includegraphics[width=1.0\columnwidth]{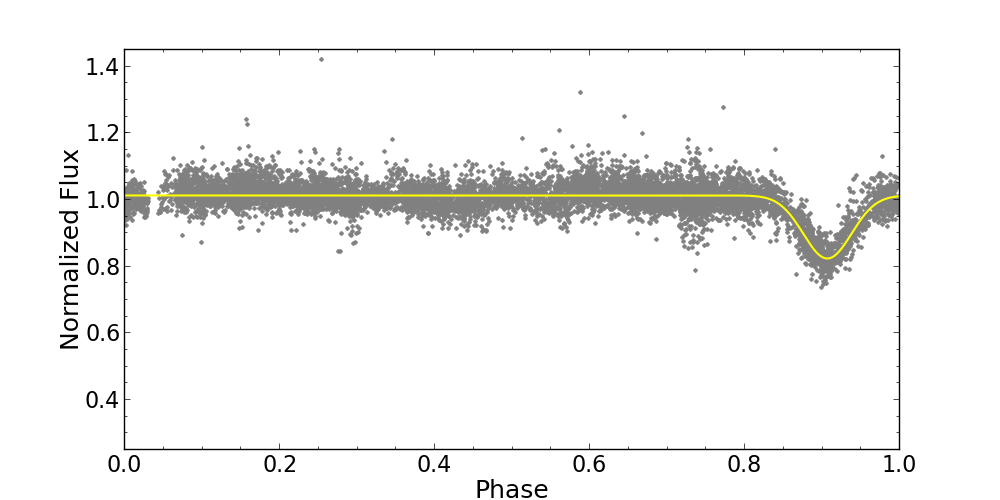}
\includegraphics[width=1.0\columnwidth]{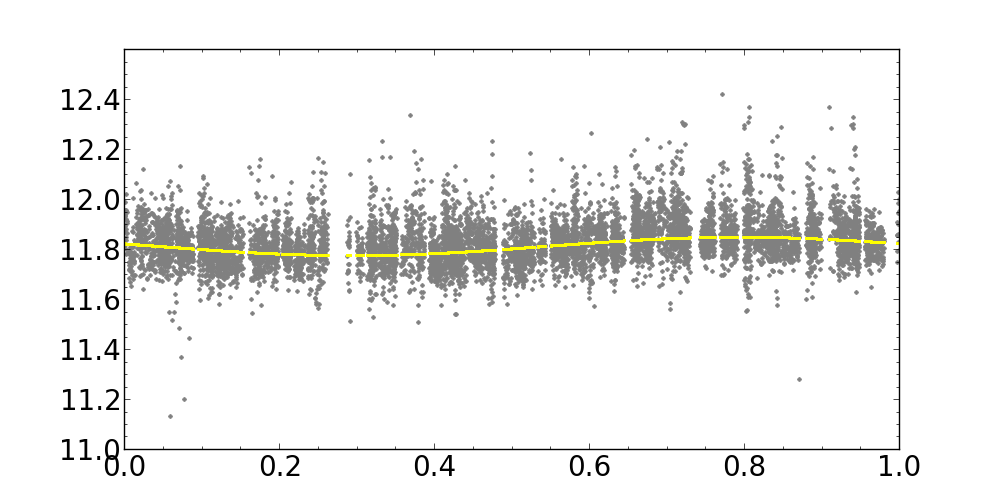}

\caption{Left: The best fit (to the 61.4905 hour eclipse in Figure \ref{fig:example_1}) to measure the depth. Gray points are two minute cadence, red points are binned in phase, yellow is the best Gaussian fit. Right: The best fit (to the 219.8386 hour variable star in Figure \ref{fig:example_1}) to measure the amplitude. Gray points are two minute cadence, red points are binned in phase, yellow is the best LS fit.}
\label{fig:example_1_fit}
\end{figure*}

\begin{figure}[!h]
\figurenum{22}
\epsscale{1}
\includegraphics[width=1.0\columnwidth]{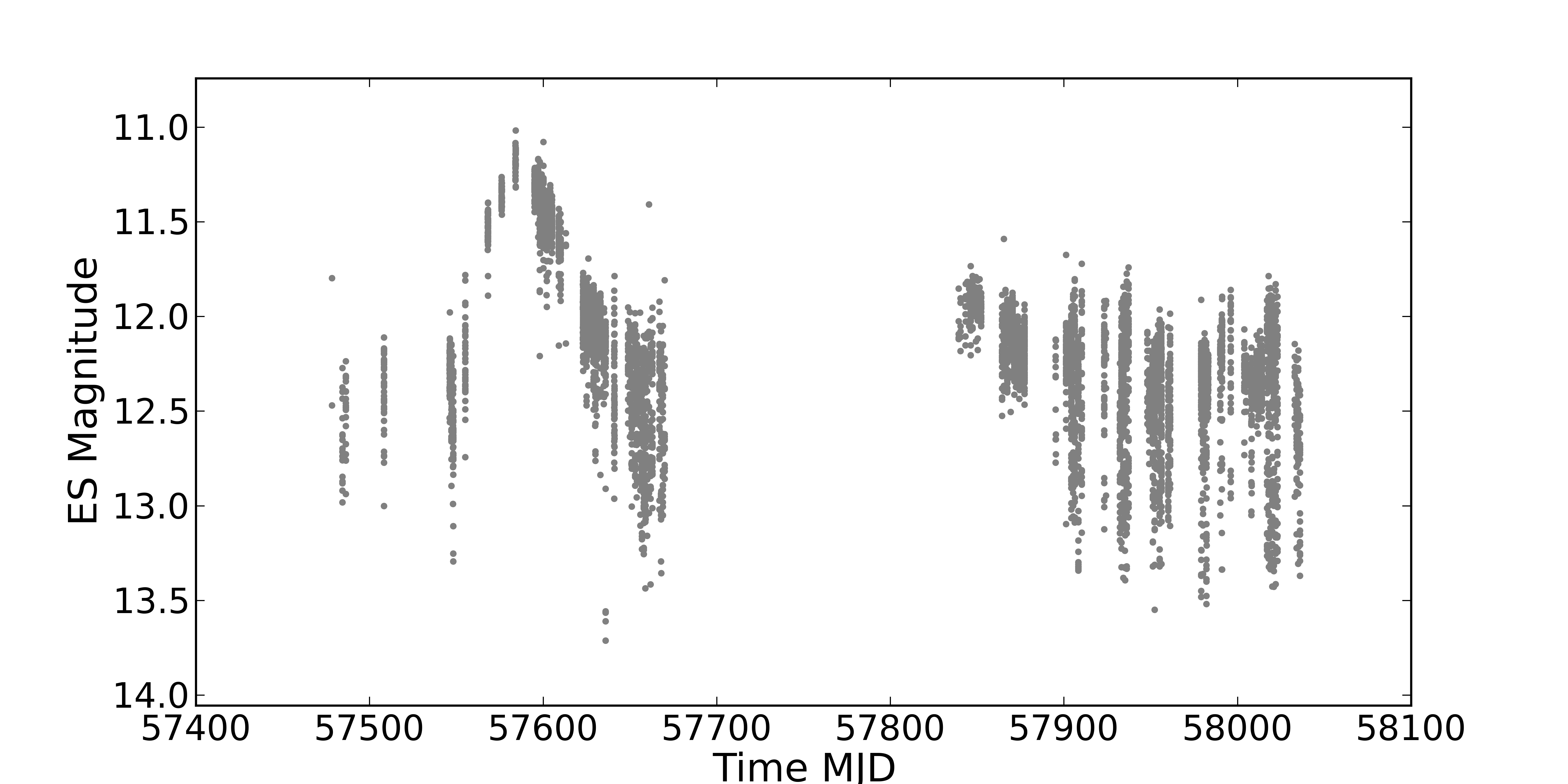}
\caption{A transient discovery with $\sim$ 100 day duration and 1.5 magnitude increase. Other long-period variables and transients including supernovae, novae, and microlensing events are detectable with the Evryscope.}
\label{fig:example_transient_variable}
\end{figure}

\clearpage
\bibliographystyle{apj}
\bibliography{ratzloff_refs}

\appendix
\section{Polar alignment procedure for an extremely-wide-field telescope}
\label{sec:polar_align}
The Evryscope's extremely-wide field of view precludes the use of a pointing/tracking model, because a conventional model optimizes the performance at the sky position at which the telescope is pointing, at the expense of the sky areas away from that direction. The Evryscope effectively points every direction simultaneously, and so the system's polar alignment accuracy is critical for the tracking performance. Conventional polar alignment strategies are made difficult because of the large pixel scale and lack of ability to point individual cameras in a wide variety of positions.

We instead developed a polar-alignment procedure that takes advantage of the Evryscope's extremely wide field of view to produce rapid sub-arcminute-precision alignment. The procedure uses the polar-facing camera to measure both the axis of rotation of the Earth and the axis of rotation of the telescope mount. Iteratively moving the telescope axis then brings the two into alignment; both axes can be measured to within a few-pixel precision. We perform the alignment as follows:

\begin{enumerate}
    \item{Measure the Earth's axis of rotation on the pole-facing camera by taking a long-exposure image with tracking turned off (10-15 minutes). The Earth's rotation axis position is measured in image coordinates using the center of the star trails. The longer the exposure, the greater the achieved positioning accuracy.}
    \item{Measure the mount's axis of rotation by taking a short-exposure image with the mount moving rapidly ($\sim$ greater than 20X tracking rate). The motion of the stars is then dominated by the mount rotation, and the center of the star trails is approximately the center of rotation of the mount (with a small offset from the Earth's rotation during the exposure).}
    \item{Iterate on the mount's polar alignment settings to bring the mount rotation axis closer to the Earth's rotation axis. It is sufficient to follow the improvements simply in pixel coordinates on the polar-facing camera. As the axes align, the offset induced by the residual Earth rotation during the mount axis alignment tends to zero, and so the mount's alignment tends to the correct position.}
\end{enumerate}

We found that this procedure could be completed in less than two hours with sub-arcminute-level alignment. This alignment procedure aligns the mount's polar axis but does not precisely locate the celestial pole in the center of the polar camera's FoV; this can be performed later by simply adjusting the mushroom pointing direction.

\newpage
\section{List of all variable discoveries}
\begin{table*}[ht]
\caption{Variable Star discoveries}
\centering
\begin{tabular}{lllllllllll}
\hline
\text{ESID} & \text{APASS ID} & \text{RA} & \text{Dec} & \text{$M_{v}$} & \text{RPM} & \text{B-V} & \text{size} & \text{spec} & \text{period} & \text{amplitude} \\ [0.1ex]
\text{ } & \text{ } & \text{ } & \text{ } & \text{ } & \text{ } & \text{ } & \text{ } & \text{ } & \text{(hours)} & \text{(delta mag)} \\ [0.1ex]
\hline

EVRJ013131.44+061855.1	&	16891108	&	22.8810	&	6.3153	&	12.99	&	11.22	&	0.97	&	ms	&	K3V	&	56.0725	&	0.047	\\
EVRJ024227.96+062556.3	&	53362	&	40.6165	&	6.4323	&	13.01	&	10.56	&	1.09	&	ms	&	K7V	&	3.3478	&	0.047	\\
EVRJ031204.99+073711.3	&	41698	&	48.0208	&	7.6198	&	13.40	&	10.22	&	1.13	&	ms	&	K7V	&	34.9678	&	0.076	\\
EVRJ031736.19+080644.3	&	34215	&	49.4008	&	8.1123	&	12.65	&	10.37	&	1.05	&	ms	&	K5V	&	10.5253	&	0.048	\\
EVRJ033741.28+064752.1	&	23523707	&	54.4220	&	6.7978	&	11.28	&	11.24	&	0.88	&	ms	&	G9V	&	4.5936	&	0.018	\\
EVRJ040342.82+051630.0	&	23508836	&	60.9284	&	5.2750	&	12.50	&	9.36	&	1.06	&	ms	&	K4V	&	30.5414	&	0.028	\\
EVRJ055815.07+082912.5	&	23801506	&	89.5628	&	8.4868	&	13.65	&	9.03	&	1.14	&	giant	&	K5	&	22.9847	&	0.047	\\
EVRJ062900.94+075330.8	&	23826908	&	97.2539	&	7.8919	&	11.86	&	9.20	&	0.85	&	ms	&	K2V	&	136.1824	&	0.053	\\
EVRJ063213.30+063835.2	&	22292962	&	98.0554	&	6.6431	&	14.31	&	9.77	&	0.86	&	ms	&	G	&	161.5992	&	0.165	\\
EVRJ064304.61+080711.6	&	22342837	&	100.7692	&	8.1199	&	11.90	&	6.36	&	1.29	&	giant	&	K	&	3.2894	&	0.050	\\
EVRJ074608.52+064450.3	&	22513221	&	116.5355	&	6.7473	&	13.81	&	9.04	&	0.96	&	ms	&	K3V	&	4.1063	&	0.068	\\
EVRJ090345.07+063356.5	&	5090425	&	135.9378	&	6.5657	&	13.27	&	10.18	&	0.85	&	ms	&	K2V	&	1001.4160	&	0.041	\\
EVRJ133939.43+080936.4	&	26935380	&	204.9143	&	8.1601	&	13.00	&	10.44	&	0.87	&	ms	&	K2V	&	3.5490	&	0.050	\\
EVRJ135123.76+074111.4	&	26926300	&	207.8490	&	7.6865	&	12.47	&	11.84	&	0.93	&	ms	&	K3V	&	103.5052	&	0.048	\\
EVRJ150518.17+062323.6	&	7678546	&	226.3257	&	6.3899	&	13.36	&	9.98	&	0.96	&	ms	&	K3V	&	4.0030	&	0.053	\\
EVRJ153240.92+054336.1	&	34080751	&	233.1705	&	5.7267	&	11.60	&	11.04	&	0.92	&	ms	&	K4V	&	29.5485	&	0.021	\\
EVRJ153936.96+061720.8	&	34088653	&	234.9040	&	6.2891	&	12.61	&	9.49	&	1.00	&	ms	&	K3V	&	1408.9650	&	0.057	\\
EVRJ155120.62+061448.8	&	34085878	&	237.8359	&	6.2469	&	13.56	&	9.81	&	1.22	&	ms	&	K6V	&	106.1767	&	0.047	\\
EVRJ155543.75+062518.8	&	34071112	&	238.9323	&	6.4219	&	11.24	&	12.09	&	1.04	&	ms	&	K4V	&	29.7894	&	0.031	\\
EVRJ164449.03+082109.7	&	34208168	&	251.2043	&	8.3527	&	13.36	&	11.12	&	0.98	&	ms	&	K5V	&	33.4419	&	0.052	\\
EVRJ173918.65+081931.4	&	34776606	&	264.8277	&	8.3254	&	13.32	&	6.90	&	0.95	&	giant	&	K	&	6.0188	&	0.013	\\
EVRJ175437.66+061028.2	&	34517257	&	268.6569	&	6.1745	&	14.08	&	11.47	&	0.88	&	ms	&	G	&	13.2881	&	0.079	\\
EVRJ180850.26+073350.4	&	34512011	&	272.2094	&	7.5640	&	13.48	&	15.00	&	1.09	&	ms	&	K2V	&	3.8693	&	0.069	\\
EVRJ182013.44+083523.6	&	34587201	&	275.0560	&	8.5899	&	12.23	&	7.27	&	1.28	&	giant	&	K	&	197.7393	&	0.040	\\
EVRJ182020.76+065445.0	&	34568159	&	275.0865	&	6.9125	&	13.25	&	9.71	&	1.10	&	ms	&	K5V	&	183.1411	&	0.063	\\
EVRJ183036.48+073707.7	&	34556082	&	277.6520	&	7.6188	&	13.19	&	10.64	&	0.86	&	ms	&	K1V	&	3.8968	&	0.077	\\
EVRJ184426.98+073442.2	&	32193828	&	281.1124	&	7.5784	&	13.44	&	10.15	&	0.87	&	ms	&	G0V	&	4.9937	&	0.133	\\
EVRJ190325.54+071516.9	&	32730341	&	285.8564	&	7.2547	&	11.47	&	9.05	&	0.98	&	ms	&	K3V	&	243.1183	&	0.017	\\
EVRJ190353.14+051812.6	&	32116381	&	285.9714	&	5.3035	&	13.33	&	9.38	&	1.04	&	ms	&	K3V	&	4.0273	&	0.052	\\
EVRJ190517.30+073520.0	&	32730666	&	286.3221	&	7.5889	&	13.62	&	12.11	&	1.30	&	ms	&	K7V	&	15.7103	&	0.041	\\
EVRJ190632.06+051345.5	&	32715501	&	286.6336	&	5.2293	&	12.64	&	9.27	&	1.19	&	giant	&	K7	&	13.0606	&	0.080	\\
EVRJ191341.81+070205.6	&	32721487	&	288.4242	&	7.0349	&	13.32	&	11.73	&	1.09	&	ms	&	K3V	&	12.3459	&	0.014	\\
EVRJ191731.06+070124.6	&	32722226	&	289.3794	&	7.0235	&	12.46	&	11.20	&	1.03	&	ms	&	K4V	&	219.8386	&	0.037	\\
EVRJ191757.24+090428.2	&	32747699	&	289.4885	&	9.0745	&	14.30	&	11.60	&	0.87	&	ms	&	G	&	22.4846	&	0.070	\\
EVRJ191908.38+083523.6	&	32746157	&	289.7849	&	8.5899	&	14.37	&	11.70	&	0.95	&	ms	&	K	&	136.0364	&	0.059	\\
EVRJ193728.03+054802.2	&	32326983	&	294.3668	&	5.8006	&	13.09	&	9.30	&	0.86	&	ms	&	G9V	&	16.6612	&	0.034	\\
EVRJ194947.38+060847.8	&	32478891	&	297.4474	&	6.1466	&	12.92	&	11.65	&	0.88	&	ms	&	K2V	&	5.8423	&	0.123	\\
EVRJ195419.58+084303.0	&	32521135	&	298.5816	&	8.7175	&	13.75	&	7.11	&	1.14	&	giant	&	K	&	236.7087	&	0.022	\\
EVRJ195728.85+074311.6	&	32498119	&	299.3702	&	7.7199	&	14.01	&	10.39	&	1.27	&	ms	&	K6V	&	4.5788	&	0.033	\\
EVRJ201533.41+082530.4	&	31613658	&	303.8892	&	8.4251	&	12.64	&	13.07	&	0.94	&	ms	&	K4V	&	28.9305	&	0.052	\\
EVRJ203320.59+090539.8	&	9498741	&	308.3358	&	9.0944	&	14.00	&	10.32	&	0.91	&	ms	&	K2V	&	3.1976	&	0.105	\\
EVRJ204952.97+054416.1	&	9315264	&	312.4707	&	5.7378	&	12.97	&	9.25	&	0.91	&	ms	&	K3V	&	161.5235	&	0.088	\\
EVRJ210125.78+082428.8	&	9339138	&	315.3574	&	8.4080	&	--	&	--	&	---	&	none	&	none	&	20.9755	&	0.018	\\
EVRJ211939.26+065648.5	&	9353342	&	319.9136	&	6.9468	&	12.90	&	9.18	&	1.09	&	ms	&	K3V	&	28.1581	&	0.038	\\
EVRJ230853.71+071107.1	&	17248213	&	347.2238	&	7.1853	&	13.14	&	11.38	&	1.09	&	ms	&	K6V	&	12.5262	&	0.139	\\

\hline

\end{tabular}
\caption{Columns 1-5 are identification numbers, right ascension and declination, and magnitude. Columns 6-9 are the reduced proper motion (RPM) and color difference (B-V) which we use to estimate the star size and spectral type (see Section 4.2.1). Columns 10 and 11 are the period found in hours, and the amplitude of the variability in magnitudes.}
\end{table*}

\clearpage
\newpage

\begin{table*}[ht]
\caption{Transient discovery}
\centering
\begin{tabular}{lllllllllll}
\hline
\text{ESID} & \text{APASS ID} & \text{RA} & \text{Dec} & \text{$M_{v}$} & \text{RPM} & \text{B-V} & \text{size} & \text{spec} & \text{duration} & \text{amplitude} \\ [0.1ex]
\text{ } & \text{ } & \text{ } & \text{ } & \text{ } & \text{ } & \text{ } & \text{ } & \text{ } & \text{(days)} & \text{(delta mag)} \\ [0.1ex]
\hline
EVRJ194754.19+073408.0 & 32510284 & 296.9758 & 7.5689 & 14.040 & 10.895 & 1.450 & ms & M1V & 100 & 1.5 \\
\hline
\end{tabular}
\end{table*}

\begin{table*}[ht]
\caption{Eclipsing Binary discoveries}
\centering
\begin{tabular}{lllllllllll}
\hline
\text{ESID} & \text{APASS ID} & \text{RA} & \text{Dec} & \text{$M_{v}$} & \text{RPM} & \text{B-V} & \text{size} & \text{spec} & \text{period} & \text{depth} \\ [0.1ex]
\text{ } & \text{ } & \text{ } & \text{ } & \text{ } & \text{ } & \text{ } & \text{ } & \text{ } & \text{(hours)} & \text{(fractional)} \\ [0.1ex]

\hline

EVRJ054324.82+070043.6	&	24006556	&	85.8534	&	7.0121	&	14.42	&	8.78	&	1.07	&	ms	&	K	&	12.3630	&	0.415	\\
EVRJ062259.52+050915.8	&	23805977	&	95.7480	&	5.1544	&	14.14	&	9.80	&	1.28	&	ms	&	M0.5V	&	159.7402	&	0.286	\\
EVRJ111947.62+085811.6	&	27552269	&	169.9484	&	8.9699	&	14.02	&	10.41	&	0.97	&	ms	&	K3V	&	56.7865	&	0.385	\\
EVRJ171609.43+070050.0	&	33836552	&	259.0393	&	7.0139	&	12.75	&	10.00	&	0.93	&	ms	&	K3V	&	16.0351	&	0.236	\\
EVRJ180755.37+063452.0	&	34507331	&	271.9807	&	6.5811	&	14.12	&	7.27	&	1.20	&	giant	&	K	&	51.7911	&	0.111	\\
EVRJ181019.32+083846.3	&	34654251	&	272.5805	&	8.6462	&	14.23	&	7.54	&	1.14	&	giant	&	A1	&	32.6179	&	0.166	\\
EVRJ181348.53+071553.6	&	34574081	&	273.4522	&	7.2649	&	13.87	&	11.70	&	0.89	&	ms	&	G	&	16.4074	&	0.145	\\
EVRJ182614.59+053454.1	&	34537571	&	276.5608	&	5.5817	&	13.21	&	10.93	&	1.15	&	ms	&	K	&	19.6076	&	0.160	\\
EVRJ191419.87+083226.5	&	32745386	&	288.5828	&	8.5407	&	14.23	&	9.78	&	0.95	&	ms	&	K	&	61.4905	&	0.189	\\
EVRJ192207.27+084849.7	&	32743749	&	290.5303	&	8.8138	&	14.15	&	9.82	&	0.98	&	ms	&	K4V	&	25.9350	&	0.196	\\
EVRJ194419.61+072333.4	&	32508956	&	296.0817	&	7.3926	&	--	&	--	&	--	&	--	&	--	&	18.5312	&	0.279	\\
EVRJ201131.20+061020.6	&	31583110	&	302.8800	&	6.1724	&	14.01	&	9.45	&	1.07	&	ms	&	K4V	&	15.1673	&	0.224	\\
EVRJ201329.93+050717.0	&	31577212	&	303.3747	&	5.1214	&	11.81	&	9.97	&	0.88	&	ms	&	G7V	&	213.0682	&	0.093	\\
EVRJ202807.01+053621.2	&	31532342	&	307.0292	&	5.6059	&	14.29	&	11.00	&	1.12	&	ms	&	K5V	&	28.2712	&	0.094	\\

\hline
\end{tabular}
\caption{Columns 1-5 are identification numbers, right ascension and declination, and magnitude. Columns 6-9 are the reduced proper motion (RPM) and color difference (B-V) which we use to estimate the star size and spectral type (see Section 4.2.1). Columns 10 and 11 are the period found in hours, and the fractional eclipse depth from normalized flux.}
\end{table*}

\newpage
\begin{figure*}[ht]
\figurenum{23}
\epsscale{1}
\includegraphics[width=.95\textwidth]{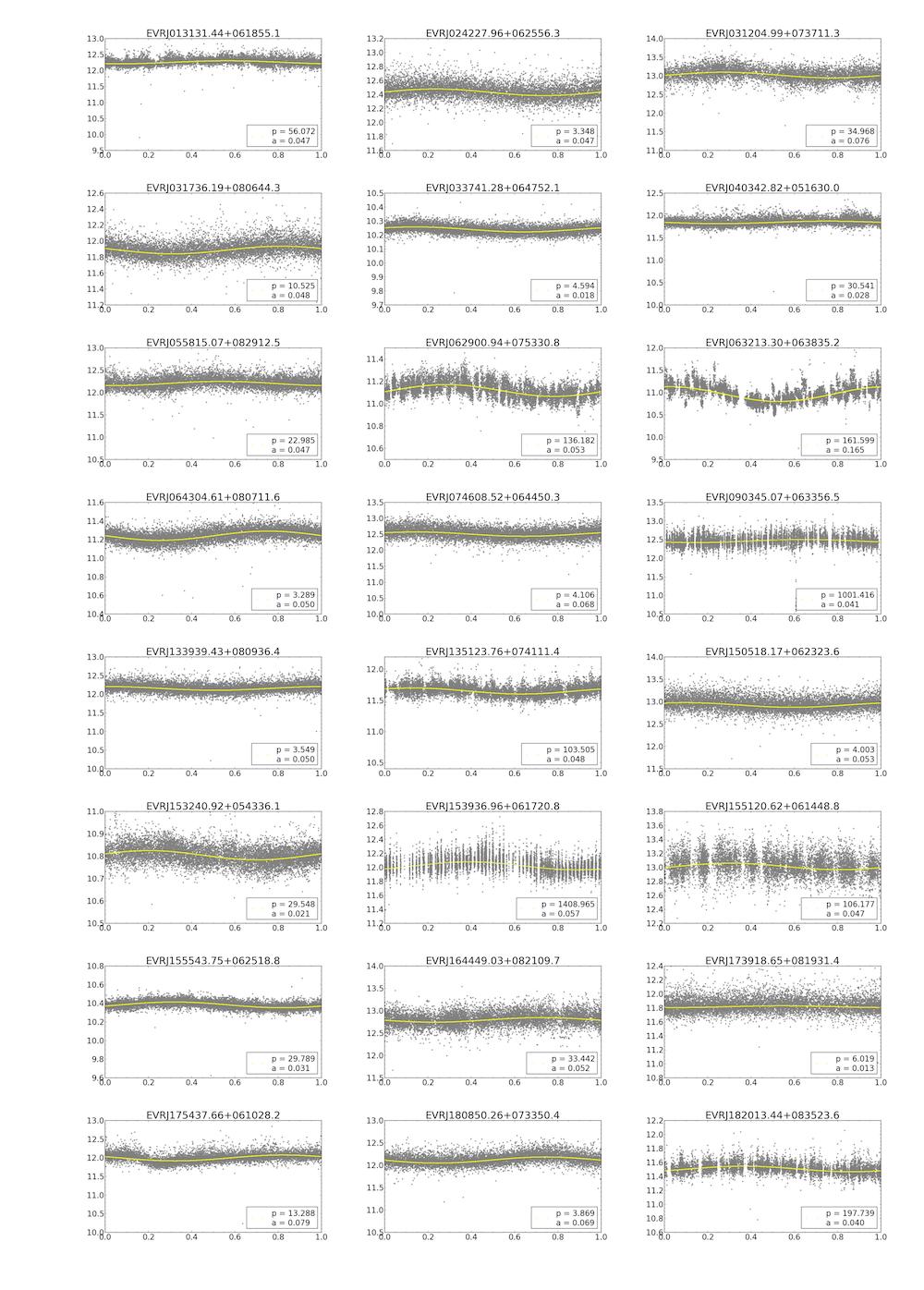}
\caption{Variable star discoveries. Y-axis is instrument magnitude, x-axis is the phase, p = period found in hours, a = amplitude change in magnitude. Gray points are two minute cadence, yellow is the best LS fit.}
\end{figure*}

\newpage
\begin{figure*}[ht]
\figurenum{23}
\epsscale{1}
\includegraphics[width=.95\textwidth]{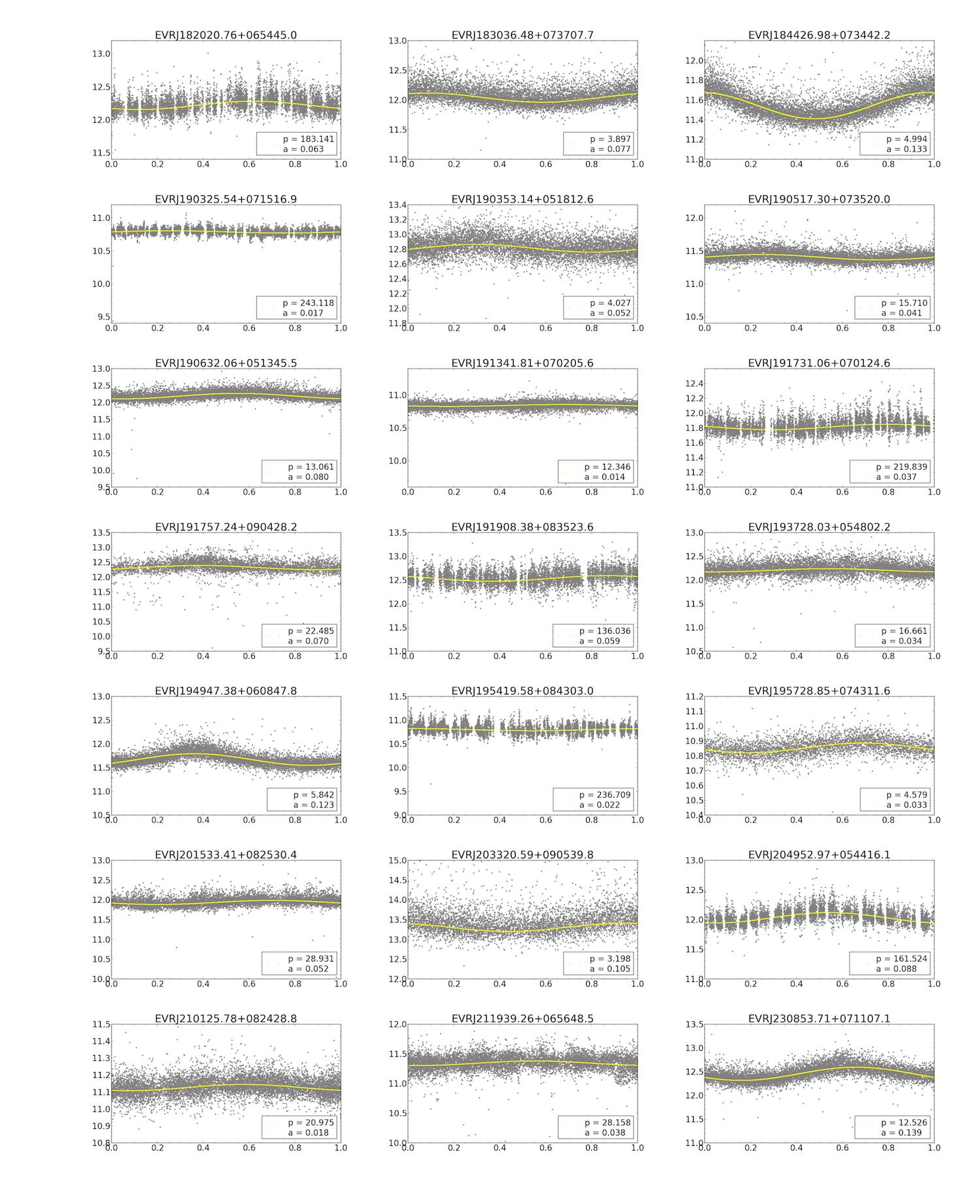}
\caption{Variable star discoveries (continued). Y-axis is instrument magnitude, x-axis is the phase, p = period found in hours, a = amplitude change in magnitude. Gray points are two minute cadence, yellow is the best LS fit.}
\end{figure*}

\newpage
\begin{figure*}[ht]
\figurenum{24}
\epsscale{1}
\includegraphics[width=.95\textwidth]{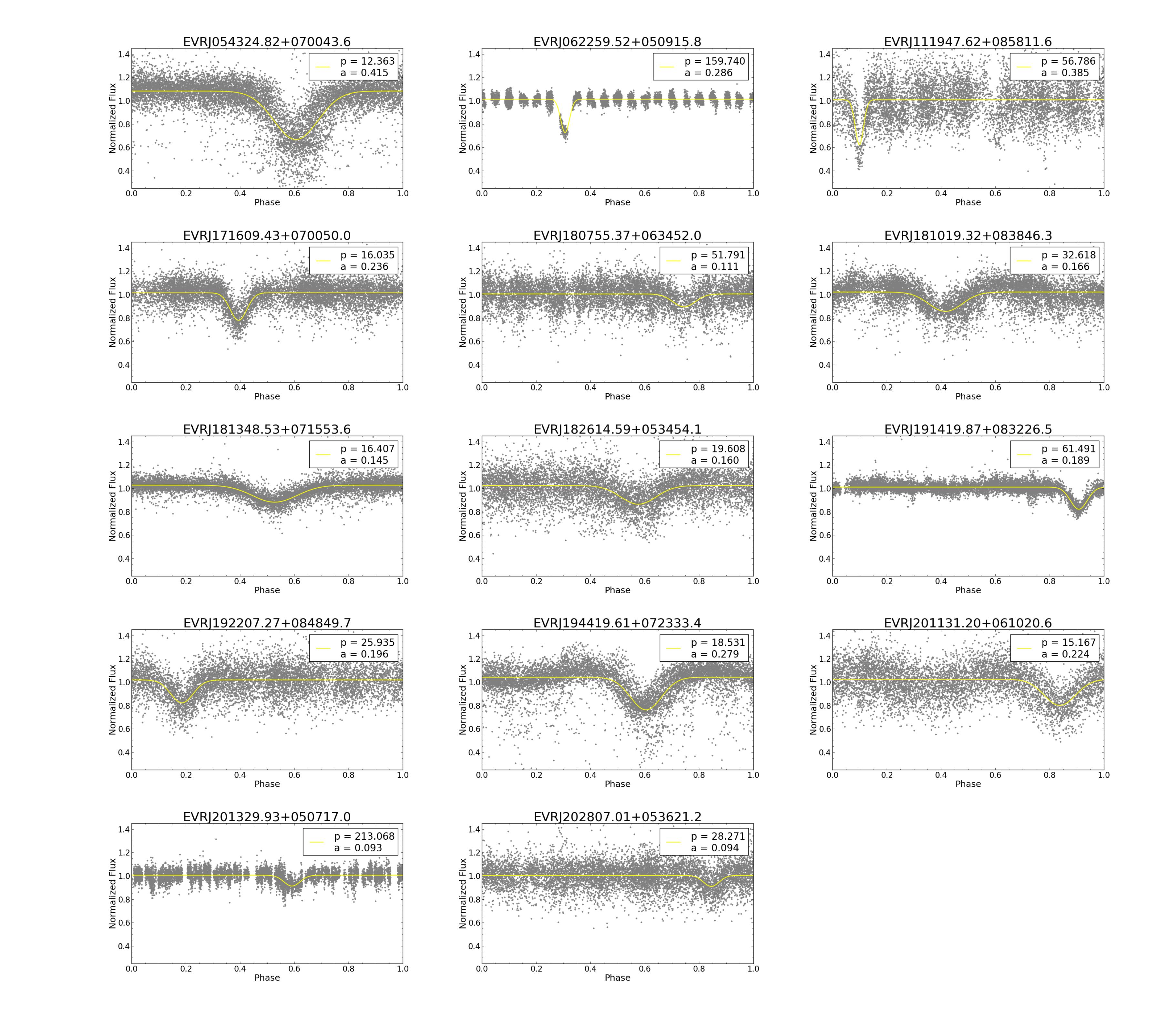}
\caption{Eclipsing Binary discoveries. Y-axis is normalized flux, x-axis is the phase, p = period found in hours, a = eclipse depth. Gray points are two minute cadence, yellow is the best fit.}
\end{figure*}

\end{document}